\documentclass[preprint,aps,prd,epsf,showpacs,amsmath,amssymb,graphics]{revtex4-1}
\usepackage{graphicx}

\begin{document}

\title{Geodesic structure of Janis-Newman-Winicour space-time}

\author{Sheng Zhou}
\author{Ruanjing Zhang }
\author{Juhua Chen} \email{jhchen@hunnu.edu.cn}
\author{Yongjiu Wang}
\affiliation{College of Physics and Information Science, Hunan
Normal University, Changsha, Hunan 410081, P. R. China}


\begin{abstract}
In the present paper we study the geodesic structure of the Janis-Newman-Winicour(JNW) space-time which contains a strong curvature naked singularity. This metric is an extension of the Schwarzschild geometry when a massless scalar field is included.
We find that the strength parameter $\mu$ of the scalar field  effects on the geodesic structure of the JNW space-time. By solving the geodesic equation and analyzing the behavior of effective potential, we investigate all geodesic types of the test particle and the photon in the JNW space-time. At the same time  we simulate  all the geodesic orbits corresponding to the energy levels of the effective potential in the JNW space-time.

\end{abstract}


\maketitle

\section{Introduction}
General Relativity has predicted many important gravitational effects, such as bending of light, precession of planetary orbits, gravitational time-delay and gravitational red-shift,  etc. The structure of geodesics helps us to understand different gravitational effects of a gravitational source. Recently the geodesics of different gravitational sources have been studied. For example, the geodesic motions in the extreme Schwarzschild-de Sitter
space-time were investigated by Podolsky \cite{Pod99}. Cruz et al. studied the geodesic structure of the Schwarzschild anti-de Sitter black hole by solving
the Hamilton-Jacobi partial differential equation \cite{0264-9381-22-6-016}. Pradhan et al. \cite{Pradhan2011474} studied the circular orbits in the extremal Reissner--Nordstr$\ddot{o}$m space-time. Pugliese et al. studied the orbits of the charged test particle in the Reissner-Nordstr$\ddot{o}$m space-time  \cite{PhysRevD.83.104052,reissner-circular} and the equatorial circular motion in the Kerr space-time \cite{PhysRevD.84.044030}. We studied the time-like geodesics of a spherically symmetric black hole in the brane-world \cite{geoBraneShengZ} and the geodesics in the Bardeen space-time \cite{geodesicBardeen}.

Both the black hole and the naked singularity are hypothetical astrophysical objects. The fact that the singularity is uncovered with an event horizon is forbidden according to Penrose's conjecture, which suggests that the cosmic censor forbids the occurrence of naked singularities. But over the past couple of decades, studies on gravitational collapse from various gravitational sources showed that the end states of complete gravitational collapse could be naked singularities \cite{PhysRevD.65.101501,Chr84,PhysRevD.19.2239,PhysRevD.38.1315,PhysRevD.76.084026,PhysRevD.58.041502,PhysRevD.47.5357,PhysRevLett.59.2137,PhysRevD.43.1416,PhysRevLett.66.994}.
So the question is that if the black hole and the naked singularity exist in nature,  there would be observational differences between them or not. Recent studies brought out some interesting characteristic differences between these objects based on the gravitational lensing and accretion disks \cite{0264-9381-27-21-215017,PhysRevD.77.124014,PhysRevD.79.043002,PhysRevD.80.024042,PhysRevD.80.104023,PhysRevD.81.104004,PhysRevD.82.124047,PhysRevD.83.024021,PhysRevD.83.104052,PhysRevD.84.044030,Pradhan2011474}.

The JNW solution is obtained as an extension of the Schwarzschild space-time when a massless scalar field is presented \cite{PhysRevLett.20.878}, which describes a spherically symmetric gravitational field that coincides with the exterior Schwarzschild solution, but the coordinate singularity in the Schwarzschild space-time becomes a naked point singularity.

The JNW line element can be written as
\begin{equation}
ds^2=-A(r)dt^2+A^{-1}(r)dr^2+B(r)d\Omega^2 ~,
\label{JNWM1}
\end{equation}
where $d\Omega^2=d\theta^2+\sin^2\theta d\phi^2$ is the
line element of a unit two-sphere, and the functions $A(r)$ and $B(r)$
are given by the following expressions
\begin{eqnarray}
  A(r) &=& \left[\frac{2r-r_0 (\mu-1)}{2r+r_0 (\mu+1)}\right]^\frac{1}{\mu} \; , \\
  B(r) &=& \frac{1}{4}\frac{[2r+r_0 (\mu+1)]^{\frac{1}{\mu}+1}}{[2r-r_0 (\mu-1)]^{\frac{1}{\mu}-1}} \; .
  \label{JNWM2}
\end{eqnarray}
The scalar field is given by
\begin{equation}
\phi =\frac{a}{\mu}\ln\left|\frac{2r-r_0(\mu-1)}{2r-r_0(\mu+1)}\right|,
\end{equation}
where $\mu$ and $a$ are linked by the relation
$\mu = \left(1+8\pi\frac{4a^2}{r_0^2} \right)$. The parameter $r_0=2m$ is related to the mass, and $\mu \in (1,\infty)$ describes the strength of the scalar field.
The minimum value of $r_{sing}=\frac{1}{2}r_0 (\mu-1)$ is a naked point singularity in the JNW space-time. When $\mu=1$, and by using the coordinate transformation $\tilde{r}=r+r_0$, the resulting metric reduces to the Schwarzschild solution. The value of the ``scalar charge'' $\mu$ corresponds to how much deviation the JNW metric is from the Schwarzschild metric.

Recently  lots of properties of the JNW space-time have been studied and the differentiating between a black hole and a singularity in the context of the JNW space-time was also discussed. For example, in Refs. \cite{patil2012acceleration} the accretion disk of the JNW naked singularity was studied. In Ref. \cite{kovacs2010can} Kovacs \textit{et al}. pointed out that an observational signature, for distinguishing rotating naked singularities from Kerr-type black holes, is that naked singularity provides a much more efficient mechanism for converting mass into radiation than black hole does. The gravitational lensing by the JNW naked singularity was studied in Refs. \cite{bozza2002gravitational,PhysRevD.65.103004,virbhadra2008time,PhysRevD.78.083004}. Liao \textit{et al}. investigated the absorption and scattering of scalar wave by the JNW naked singularity \cite{LiaCheHua14}. The circular geodesics and accretion disks in the JNW space-time have been studied previously \cite{PhysRevD.85.104031}, and the range of the parameter $\mu$ was divided into three regions $(1,2)$, ~$(2,\sqrt{5})$ ~and ~$(\sqrt{5},\infty)$ where structure of the circular geodesics is qualitatively different. In the present paper we will focus on studying all types of geodesic orbits by solving the geodesic equation and analyzing the behavior of effective potential. In the JNW space-time, for time-like geodesics, we take the viewpoint in Ref. \cite{PhysRevD.85.104031} and discuss the three kinds of geodesic structures characterized by $\mu$. For the null geodesic, we find that the range of $\mu$ can be divided into two regions $(1,2) ~and~ (2,\infty)$ which distinguish two different null geodesic structures. We plot all the possible geodesic orbits of the test particle and photon for all cases which are allowed by the energy level in the JNW space-time.

 The present paper is organized as follows: In Section \ref{section geodesics} we define the effective potential and give the condition for circular orbit. In Section \ref{Time-like geodesics} we study all cases of time-like geodesics and in Section \ref{null geodesics} we discuss the null geodesics. A brief conclusion is given in Section \ref{Summary and Conclusions}.

\section{Geodesic Equation in the JNW space-time}\label{section geodesics}
Now we turn to set up the geodesic equation in the JNW space-time by solving the Lagrange equation. For a general static spherically symmetric solution (\ref{JNWM1}), the corresponding Lagrangian reads
\begin{eqnarray}\label{eq:Lagrangian}
2\mathcal{L}=-A(r)\dot{t}^2+A(r)^{-1}\dot{r}^2+B(r)(\dot{\theta}^2+sin^2\theta \dot{\phi}^2),
\end{eqnarray}
where the dot ``." represents the derivative with respect to the affine parameter $\tau$, along the geodesic.
The equation of motion is
\begin{equation}
\dot{\Pi}_q-\frac{\partial{\mathcal {L}}}{\partial q}=0 ,
\end{equation}
where $\Pi_q=\partial {\mathcal {L}}/\partial \dot{q}$ is the momentum to coordinate $q$. Since the Lagrangian is independent of $(t,\phi)$, the corresponding conjugate momentums are conserved, therefore
\begin{equation}\label{eq:E}
\Pi_t = -A(r)\dot{t} = -E,
\end{equation}
\begin{equation}\label{eq:L}
\Pi_\phi = B(r)sin^2\theta\dot{\phi} = L,
\end{equation}
where $E$ and $L$ are motion constants.

From the motion equation of the coordinate $\theta$
\begin{eqnarray}
\dot{\Pi}_\theta-\frac{\partial{L}}{\partial \theta}=0,
\end{eqnarray}
we have
\begin{equation}\label{eq:motion for theta}
\frac{d(B(r)\dot{\theta})}{d\tau} = B(r) sin\theta cos\theta \dot{\phi}^2.
\end{equation}
If we choose the initial conditions $\theta = \pi/2$, $\dot{\theta} =\ddot{\theta}= 0$, and according to Eq. (\ref{eq:motion for theta}), the geodesic motion is restricted on the equatorial plane. So the Eq. (\ref{eq:L}) could be further simplified into
\begin{equation}\label{eq:L1}
\Pi_\phi = B(r)\dot{\phi} = L,
\end{equation}
from Eqs. (\ref{eq:E}) and (\ref{eq:L}), the Lagrangian (\ref{eq:Lagrangian}) can be written in the following form
\begin{equation}
2{\mathcal {L}}\equiv h = \frac{E^2}{A(r)}-\frac{\dot{r}^2}{A(r)}-\frac{L^2}{B(r)}.
\end{equation}
By solving the above equation, we can obtain the radial motion equation,
\begin{equation}\label{eq:motion}
\dot{r}^2 = E^2 - V_{\rm{eff}}^2,
\end{equation}
where we define $V_{\rm{eff}}^2$ as an effective potential
\begin{equation}\label{eq:effective potential}
V_{\rm{eff}}^2 = A(r)\left(h+\frac{L^2}{B(r)}\right).
\end{equation}
For circular geodesics, we have
\begin{equation}\label{circular condition}
V_{\rm{eff}}=E \;  \; \text{and} \; \frac{\partial V_{\rm{eff}}}{\partial r}=0. \;
\end{equation}
The circular orbit is stable against the small perturbations in the radial direction if the effective potential admits a minimum
\begin{equation}\label{stable condition}
\frac{\partial^2 V_{\rm{eff}}}{\partial r^2}>0,
\end{equation}
or the orbit is unstable if the effective potential admits a maximum. The inflection point of the effective potential, i.e. $\frac{\partial^2 V_{\rm{eff}}}{\partial r^2}=\frac{\partial V_{\rm{eff}}}{\partial r}=0$ corresponds to the marginally stable circular orbit.

\section{Time-like geodesics} \label{Time-like geodesics}
For the time-like geodesic $h=1$, the effective potential $V^2_{\rm{eff}}$ becomes

\begin{equation}\label{eq:time-like effective potential}
V_{\rm{eff}}^2 = A(r)(1+\frac{L^2}{r^2}).
\end{equation}

\begin{figure}
	\includegraphics[scale=0.6]{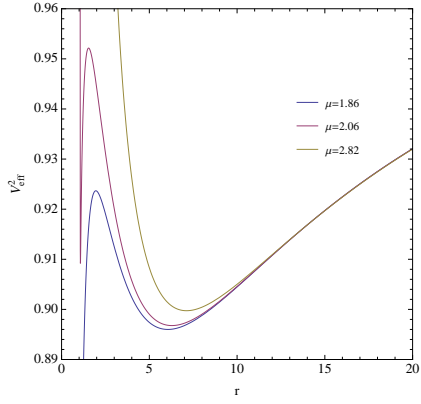}\\
	\caption{The effective potential $V^2_{\rm{eff}}$ of the radial motion is plotted as a function of radial coordinate $r$ for different values of $\mu$ in three ranges, i.e., $\mu =1.86 \in (1,2)$, $\mu =2.06 \in (2,\sqrt{5})$ and $\mu =2.82 \in (\sqrt{5},\infty)$. }\label{fig:V-time-like-mu}
\end{figure}

By imposing the conditions Eqs.(\ref{circular condition},\ref{stable condition}) into Eq.(\ref{eq:time-like effective potential}) , we get
\begin{eqnarray}
L^2=\frac{r_0}{2r-r_0}B(r), \\
E^2=\frac{2r}{2r-r_0}A(r) ,\\
4r^2-8rr_0+r_0^2(\mu^2-1)>0,
\end{eqnarray}
from the above equations we can get the radius $r=\frac{r_0}{2}$ for photon sphere, and characterize the time-like geodesics by three distinct ranges of the parameter, i.e.  $\mu \in (1,2)$, $\mu \in (2,\sqrt{5})$, $\mu \in (\sqrt{5},\infty)$, in which the geodesic structures are very different. In Fig.\ref{fig:V-time-like-mu},  three kinds of effective potential corresponding to the three ranges of the parameter $\mu$ are plotted, and we will discuss all the possible time-like geodesic orbits for these cases, respectively.

According to Eqs.(\ref{eq:motion},\ref{eq:time-like effective potential}), the motion equation of a particle reads

\begin{equation}\label{eq:time-like motion}
	\dot{r}^2 = E^2 - A(r)(1+\frac{L^2}{r^2}).
\end{equation}

By using Eq.(\ref{eq:L1}) and making a change of variable $u^{-1}=r$, we can obtain the orbit motion equation of the test particle

\begin{equation}\label{orbit equation for particle}
	(\frac{du}{d\phi})^2=\frac{u^4}{16}(\frac{E^2ab}{L^2}-4ab-\frac{a^{2-1/\mu} ~b^{2+1/\mu}}{L^2}),
\end{equation}
where $a=2u^{-1}+r_0(1-\mu)$ and $b=2u^{-1}+r_0(1+\mu)$.

Solving the above motion equation numerically for the three ranges of the parameter $\mu$, we get all types of geodesic orbits in the JNW space-time in detail.

\subsection{Case $\mu \in(1,2)$}

\begin{figure}
	\includegraphics[scale=0.6]{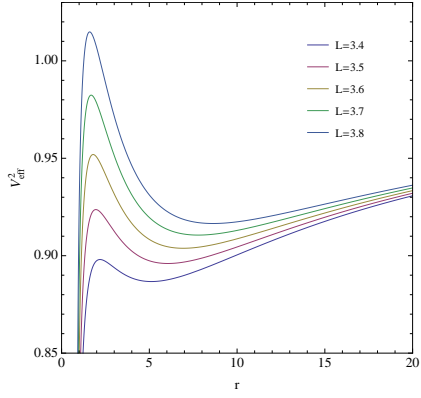}\\
	\caption{The effective potential $V^2_{\rm{eff}}$ of the radial motion is plotted as a function of radial coordinate $r$ for $R_0=2, \mu =1.86 \in (1,2)$ with different values of $L$.}\label{fig:V-time-like-mu1-L}
\end{figure}

In Fig.\ref{fig:V-time-like-mu1-L} the general behavior of the effective potential is shown as a function of the radius with a fixed value of the parameter  $\mu =1.86 \in (1,2)$ for different values of the angular momentum $L$. The effective potential has a maximum and a minimum which corresponds to the unstable and stable circular orbits, respectively.  From the effective potential, we can also expect a bound orbit, a terminating orbit and an escape orbit for the test particle.

\subsubsection{Time-like circular orbit }

\begin{figure}
	\includegraphics[scale=0.41]{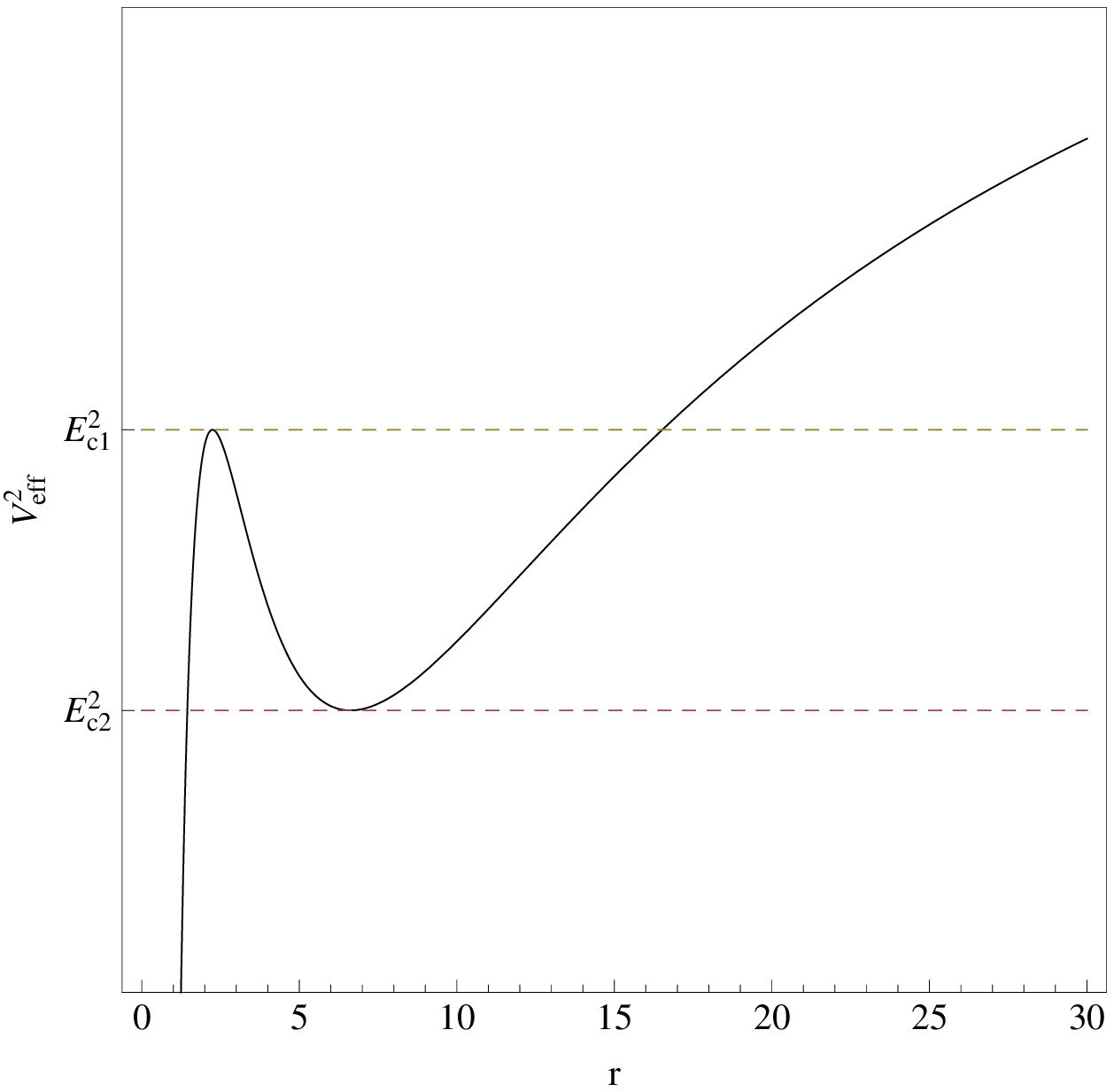}
	\includegraphics[scale=0.4]{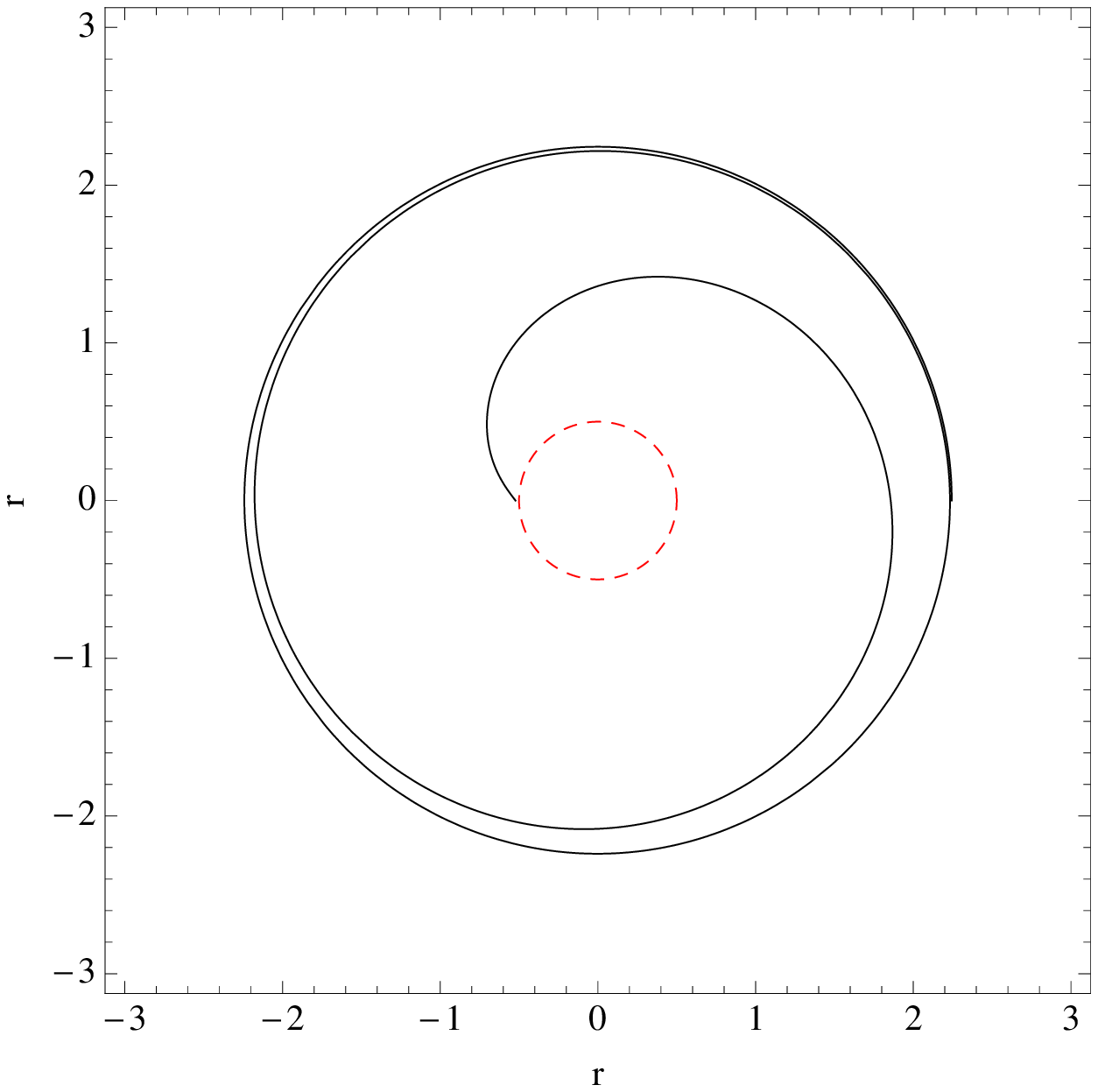}
	\includegraphics[scale=0.405]{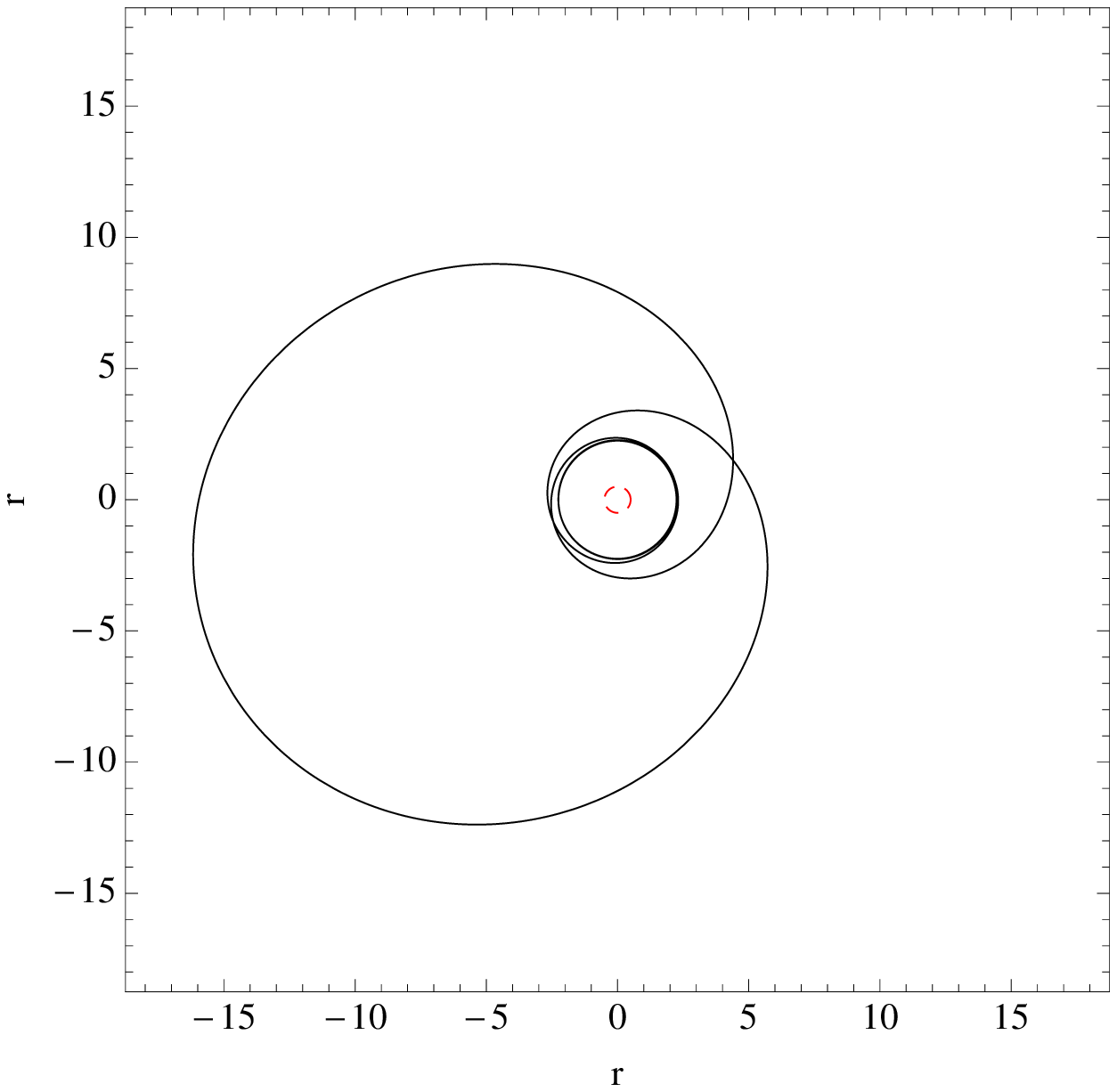}
	\caption{Examples of unstable time-like circular geodesics of JNW space-time with $E_{c1}^2 =0.925, L=3.6, r_0=2, \mu=1.5$.}\label{fig:V1 time-like circle}
\end{figure}

In Fig.\ref{fig:V1 time-like circle}, there exist stable and unstable circular orbits. When the energy of the test particle is equal to the peak value of the effective potential curve $E_{c1}$, the test particle will be on an unstable circular orbit, i.e., a tiny perturbation makes the particle fall into the singularity or move on an bound orbit when it fall into the right side of the potential barrier instead; When the energy of the particle is equal to the bottom value $E_{c2}$ of the effective potential, the particle will move on a stable circular orbit.

\subsubsection{Time-like bound orbit}

\begin{figure}
	\includegraphics[scale=0.395]{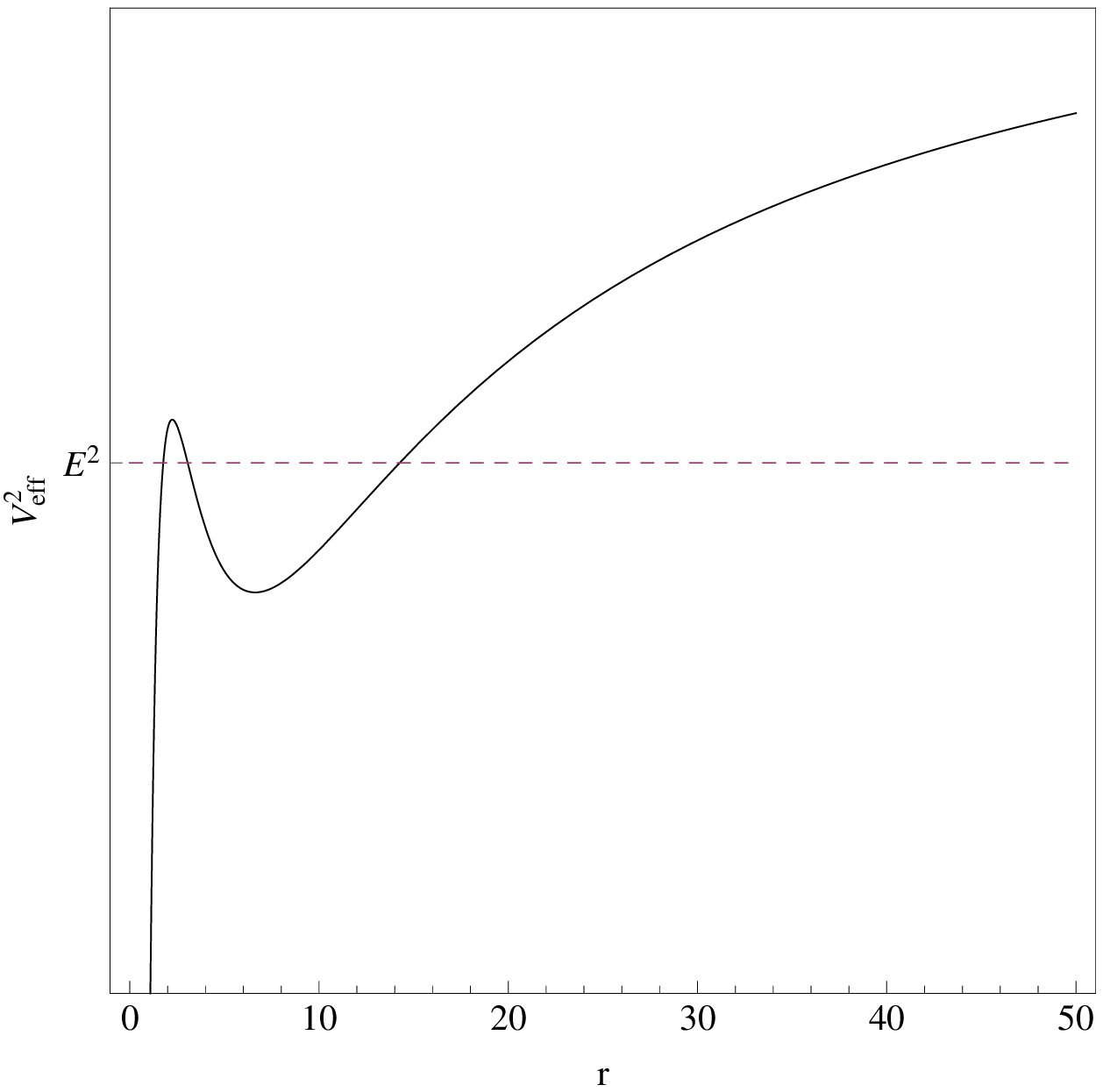}
	\includegraphics[scale=0.4]{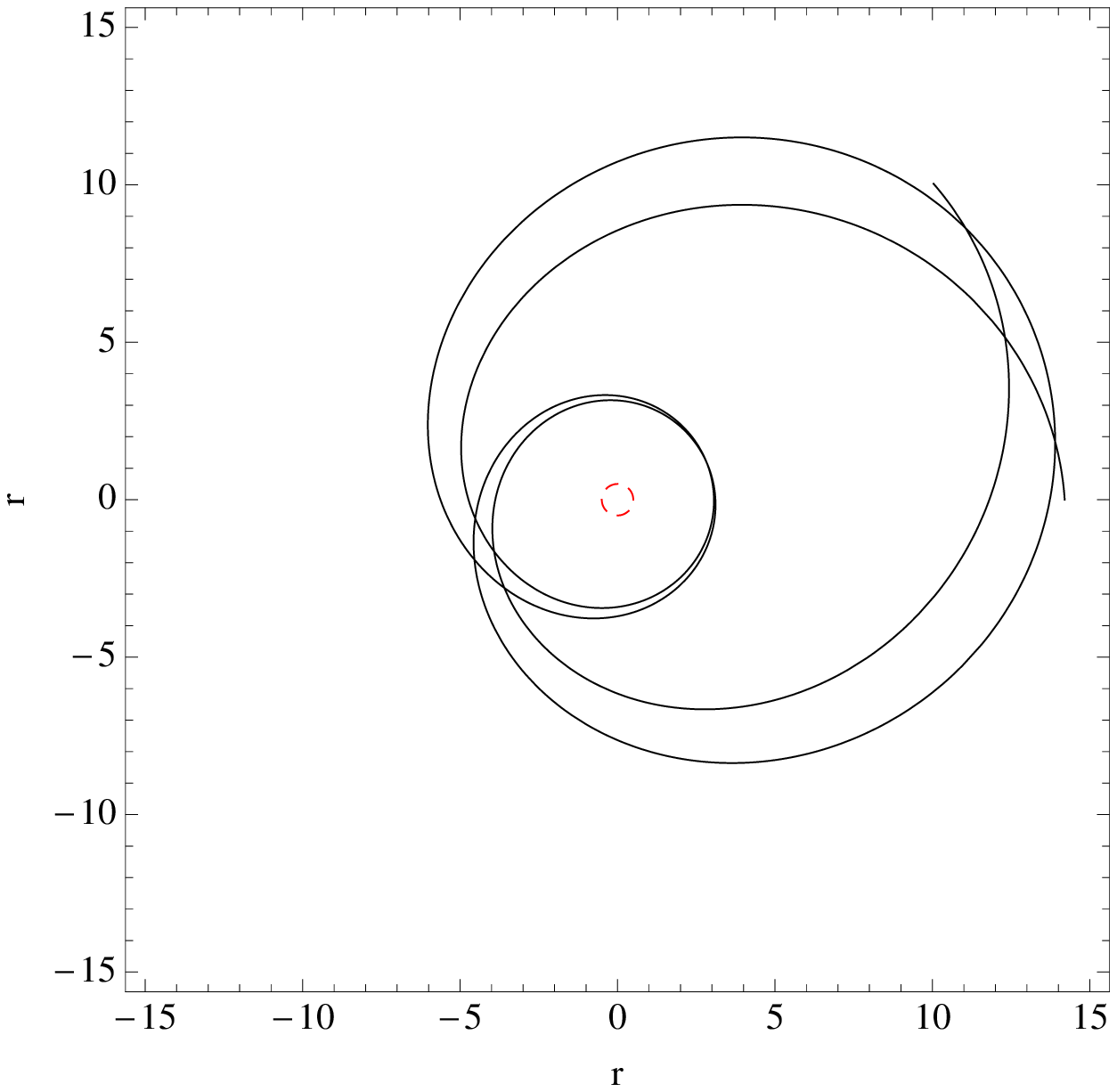}
	\caption{Example of the time-like bound orbit of the JNW space-time with $E^2 =0.92, L=3.6, r_0=2, \mu=1.5$.  }\label{fig:V1 time-like bound}
\end{figure}

The time-like bound orbit for $\mu=1.5\in(1,2)$ is plotted in Fig.\ref{fig:V1 time-like bound}. If the energy of the particle is between the peak value and the bottom value of the potential, the particle will move on a bound orbit with the radius between an aphelion and a perihelion for this case.

\subsubsection{Time-like terminating and terminating escape orbits}

\begin{figure}
	\includegraphics[scale=0.405]{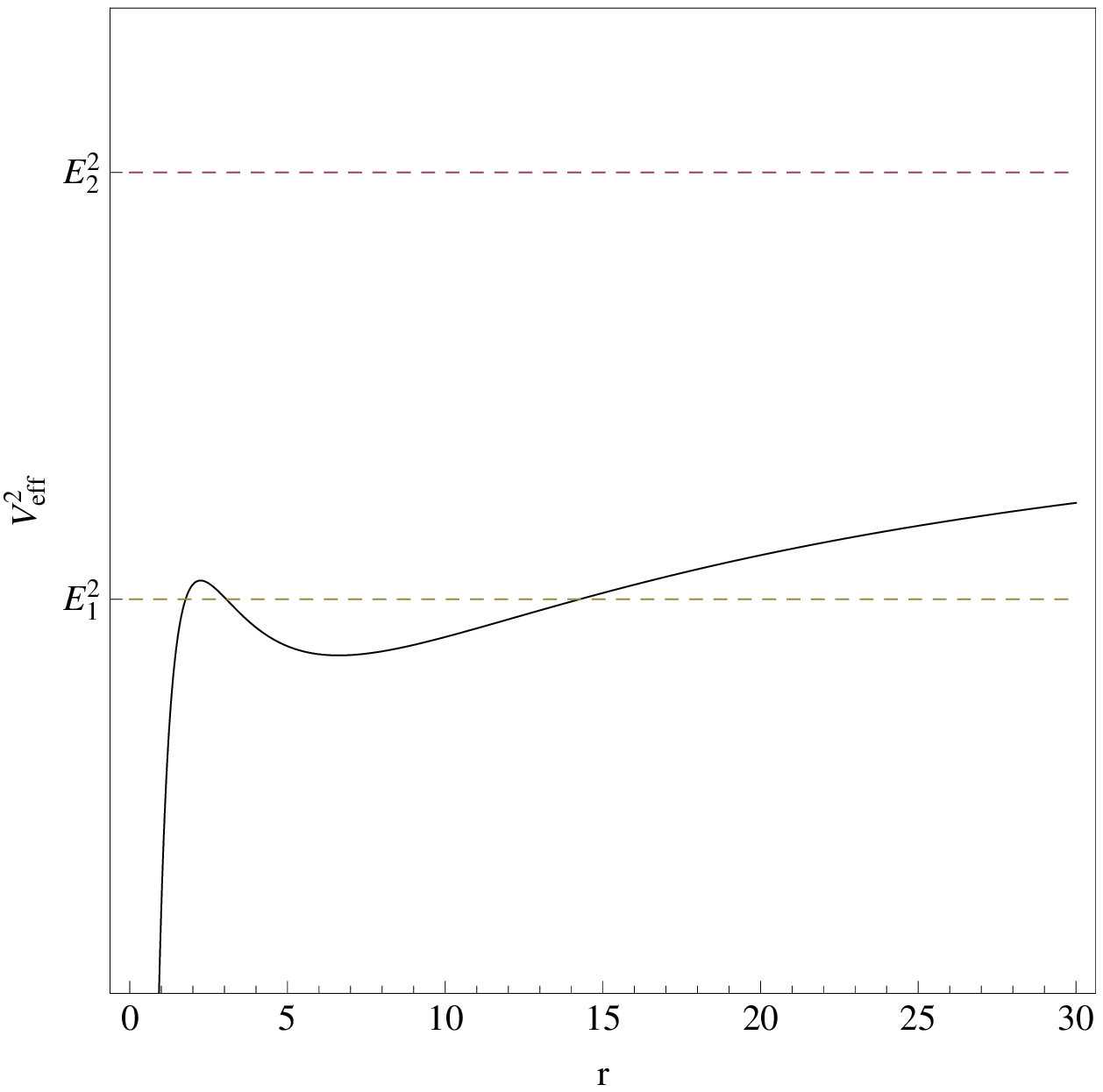}
	\includegraphics[scale=0.4]{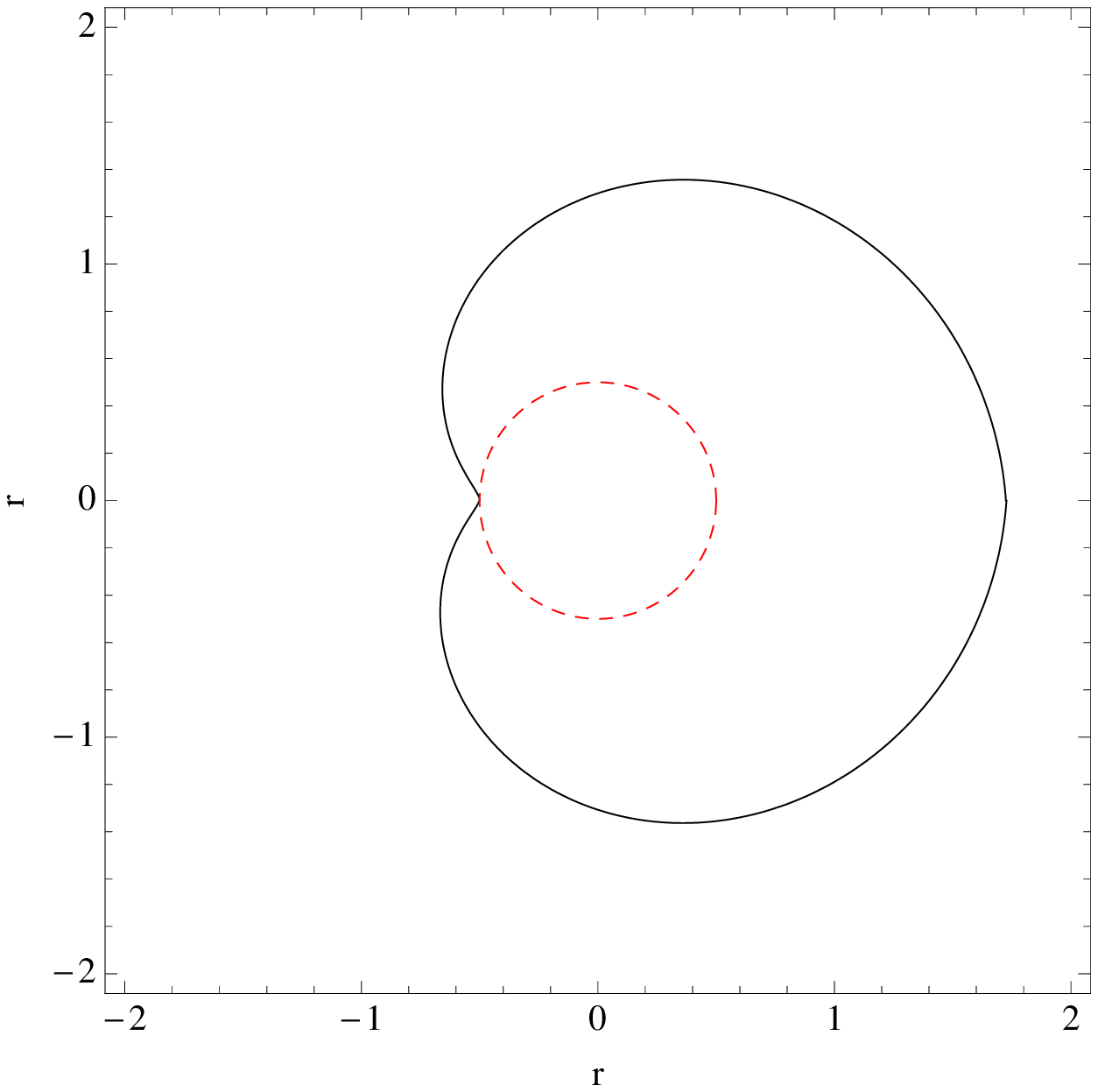}
	\includegraphics[scale=0.407]{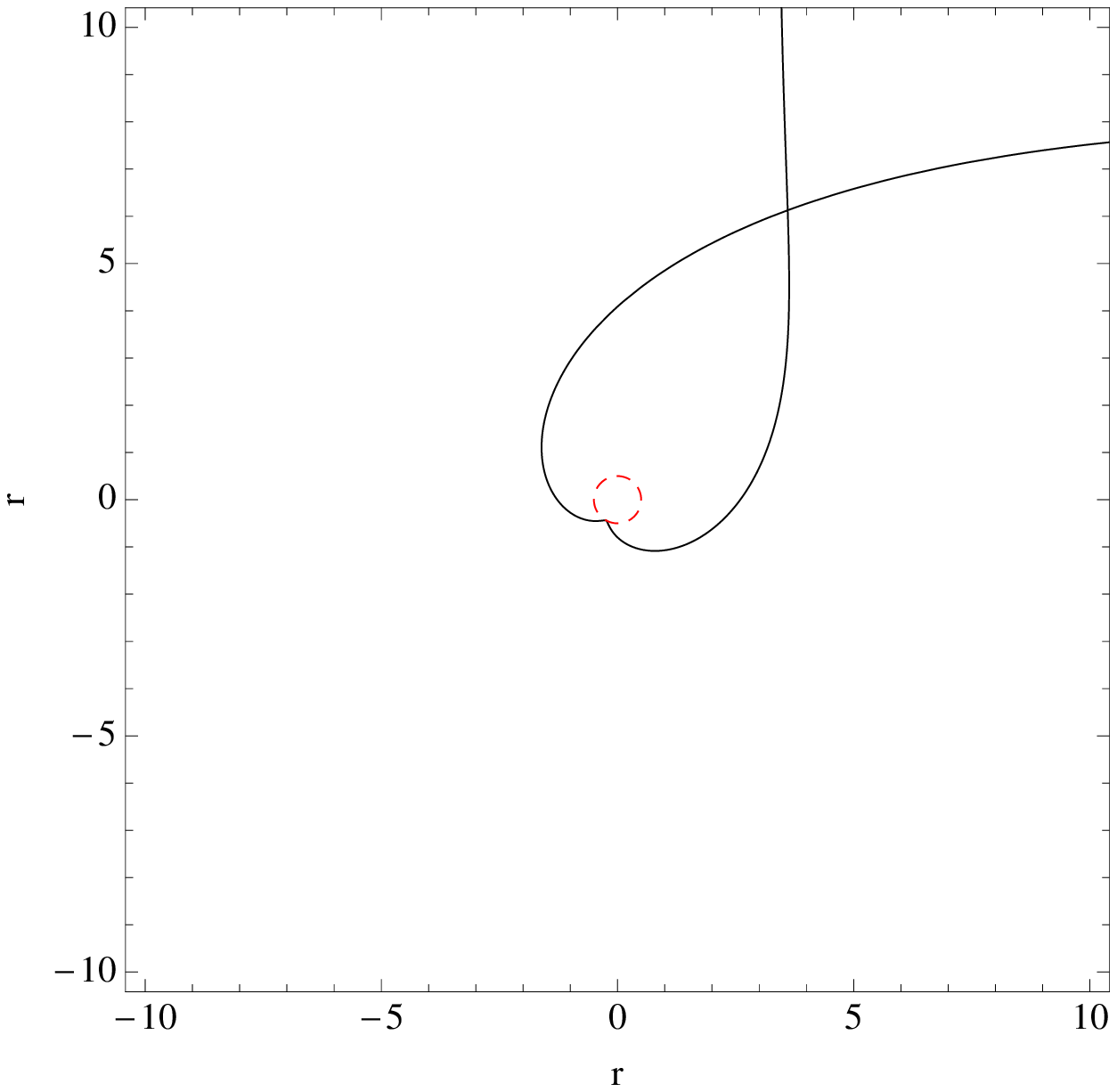}
	\caption{Examples of the time-like terminating and terminating escape orbits of the JNW space-time for $E_{1}^2 =0.92, E_{2}^2 =1.05$ with fixed $L=3.6, r_0=2, \mu=1.5$.  }\label{fig:V1 time-like terminating orbit}
\end{figure}

Fig.\ref{fig:V1 time-like terminating orbit} shows terminating and terminating escape orbits. I) When the energy of the particle is lower than the peak value of effective potential, the particle will move on a terminating orbit from a finite distance on the left side of the potential barrier and end at the singularity eventually; II) When the energy of the particle is higher than the peak value of the potential, The test particle will move on a terminating escape orbit and will end at the singularity if the particle comes from infinity.

\subsection{Case $\mu \in(2,\sqrt{5})$}

\begin{figure}
	\includegraphics[scale=0.6]{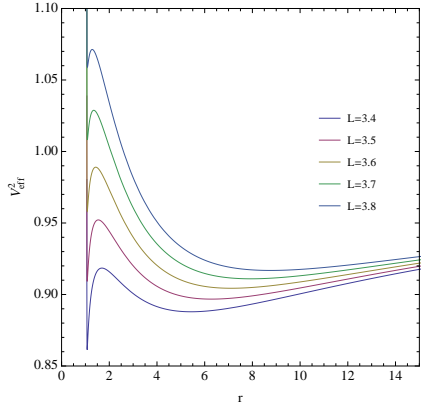}\\
	\caption{The effective potential $V^2_{\rm{eff}}$ of the radial motion is plotted as a function of radial coordinate $r$ for $\mu =2.06 \in (2,\sqrt{5})$ for different values of $L$.}\label{fig:V-time-like-mu2-L}
\end{figure}

In Fig.\ref{fig:V-time-like-mu2-L} the general behavior of the effective potential is shown as a function of the radius with a fixed value of the parameter  $\mu =2.06 \in (2,\sqrt{5})$ for different values of the angular momentum $L$. The effective potential has one maximum between two minimums, which correspond to the unstable and stable circular orbits, respectively. There also exists a bound orbit or an escape orbit, but no terminating orbit, i.e., particle will not fall into the singularity.

\subsubsection{Time-like circular orbit }

\begin{figure}
	\includegraphics[scale=0.4]{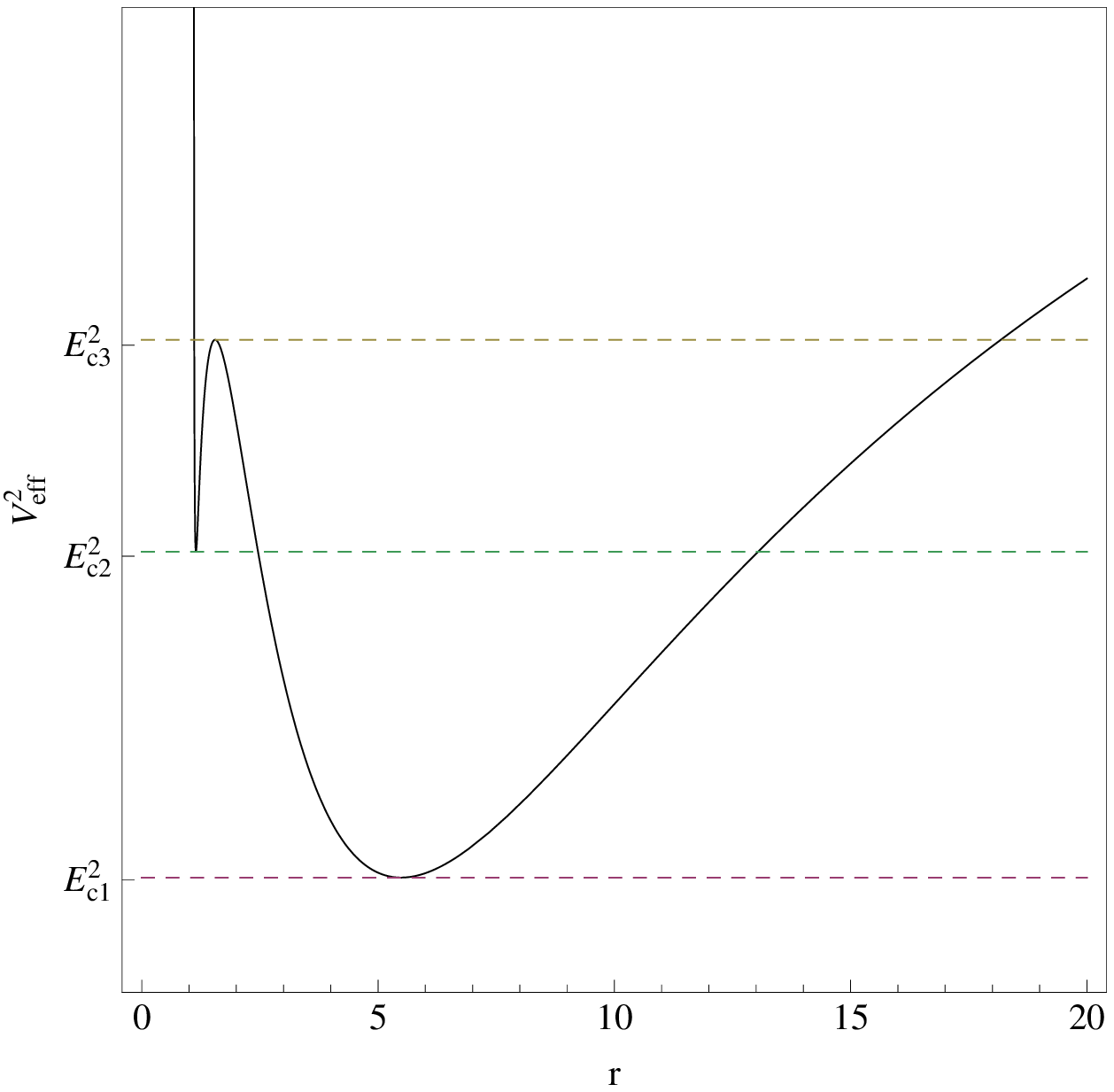}
	\includegraphics[scale=0.4]{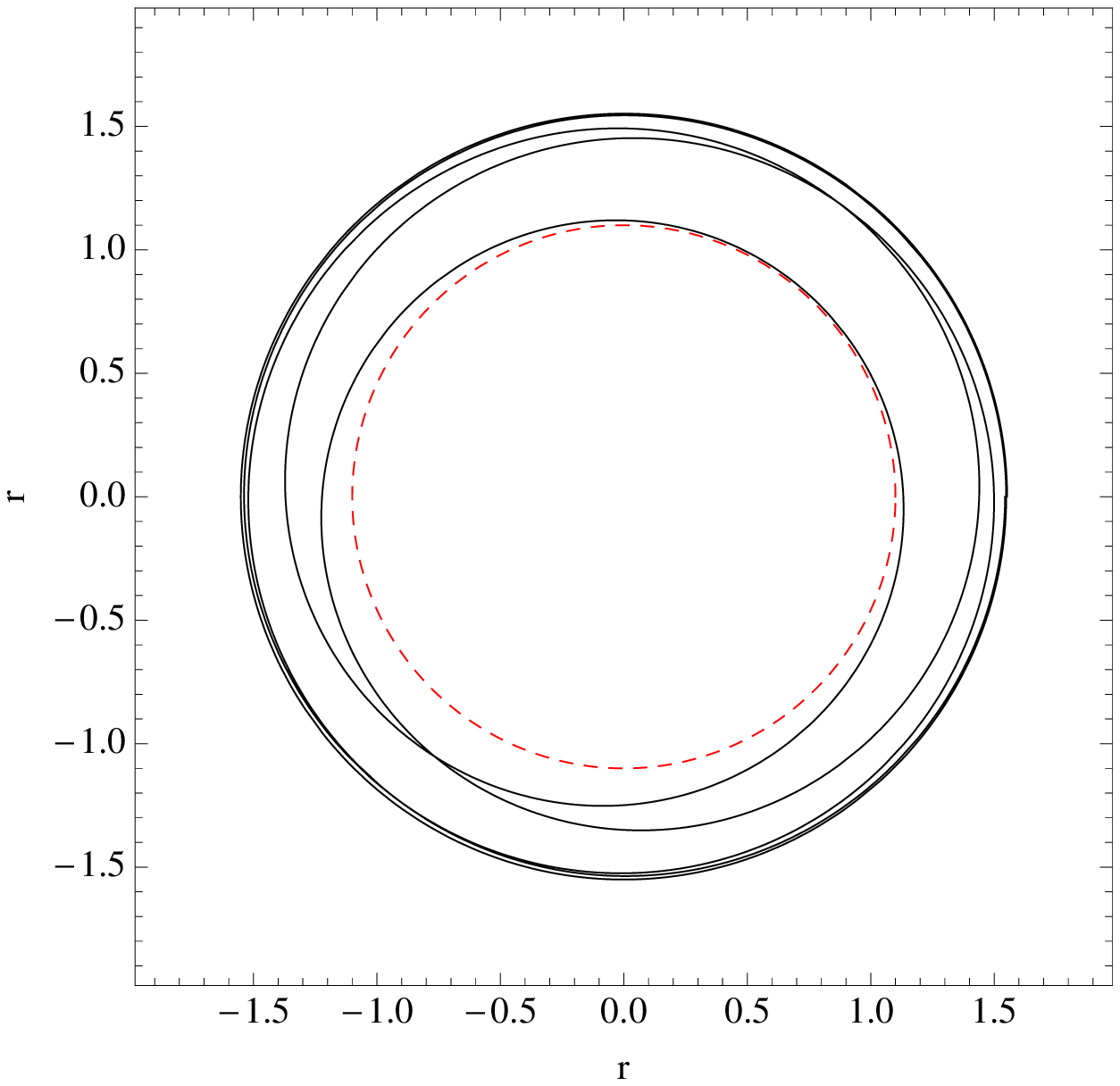}
	\includegraphics[scale=0.4]{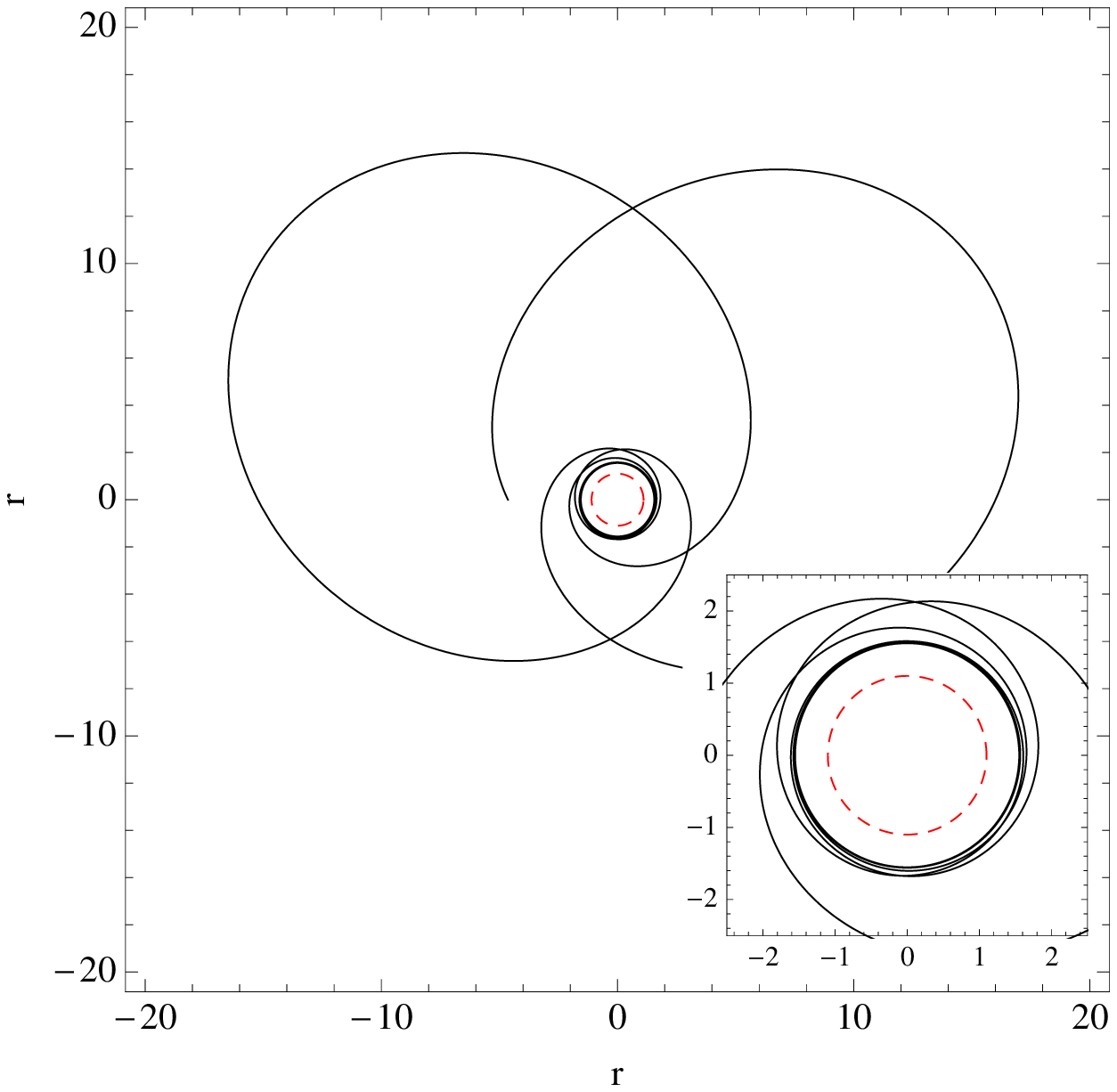}
	\caption{Examples of the time-like unstable circular orbit of the JNW space-time with $E_{c3}^2 =0.926, L=3.4, r_0=2, \mu=2.1$.  }\label{fig:V2 time-like circle}
\end{figure}

In Fig.\ref{fig:V2 time-like circle} we can see that there is an unstable circular orbit between two stable circular orbits. I) When the energy of the particle is equal to the bottom values of the effective potential curve $E_{c1}$ or $E_{c2}$, the particle will orbit on  two different stable circular orbits. II) When the energy of the particle is equal to the peak value of the potential $E_{c3}$, it will move on an unstable circular orbit. Under this case the test particle will move on two kinds of bound orbits on each side of the potential barrier due to a tiny perturbation.

\subsubsection{Time-like bound orbit}

\begin{figure}
	\includegraphics[scale=0.4]{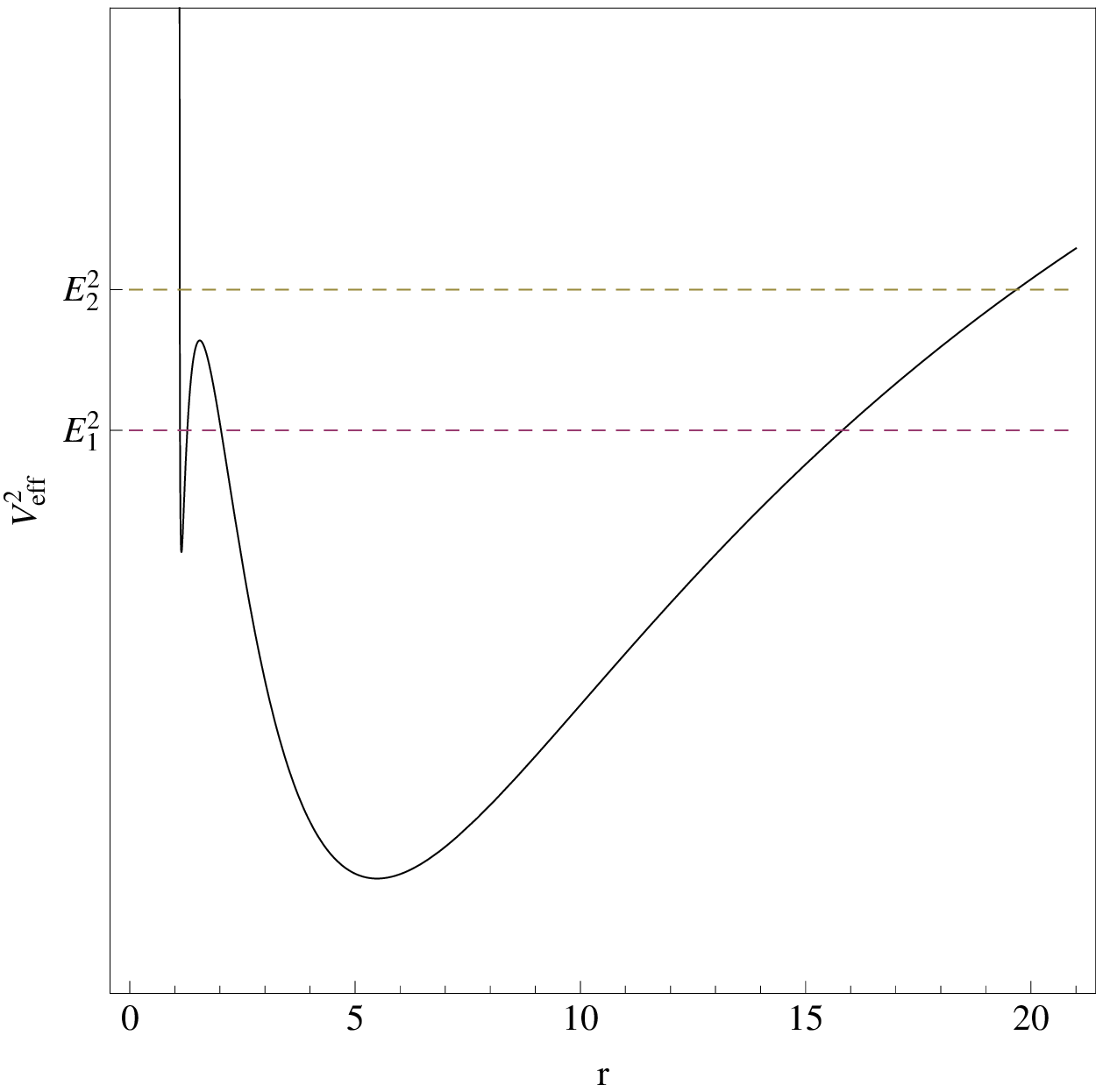}
	\includegraphics[scale=0.41]{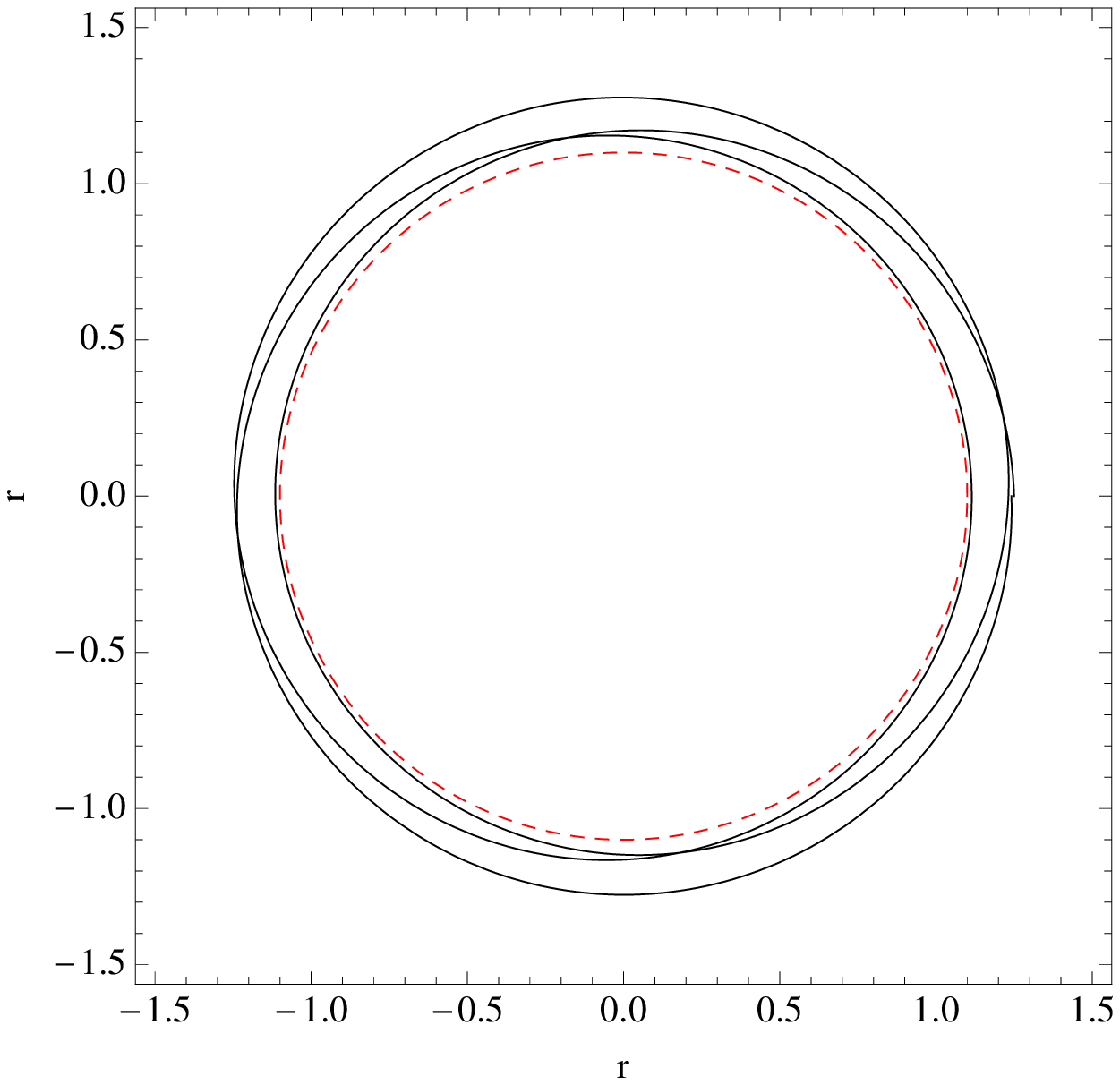}\\
	\includegraphics[scale=0.405]{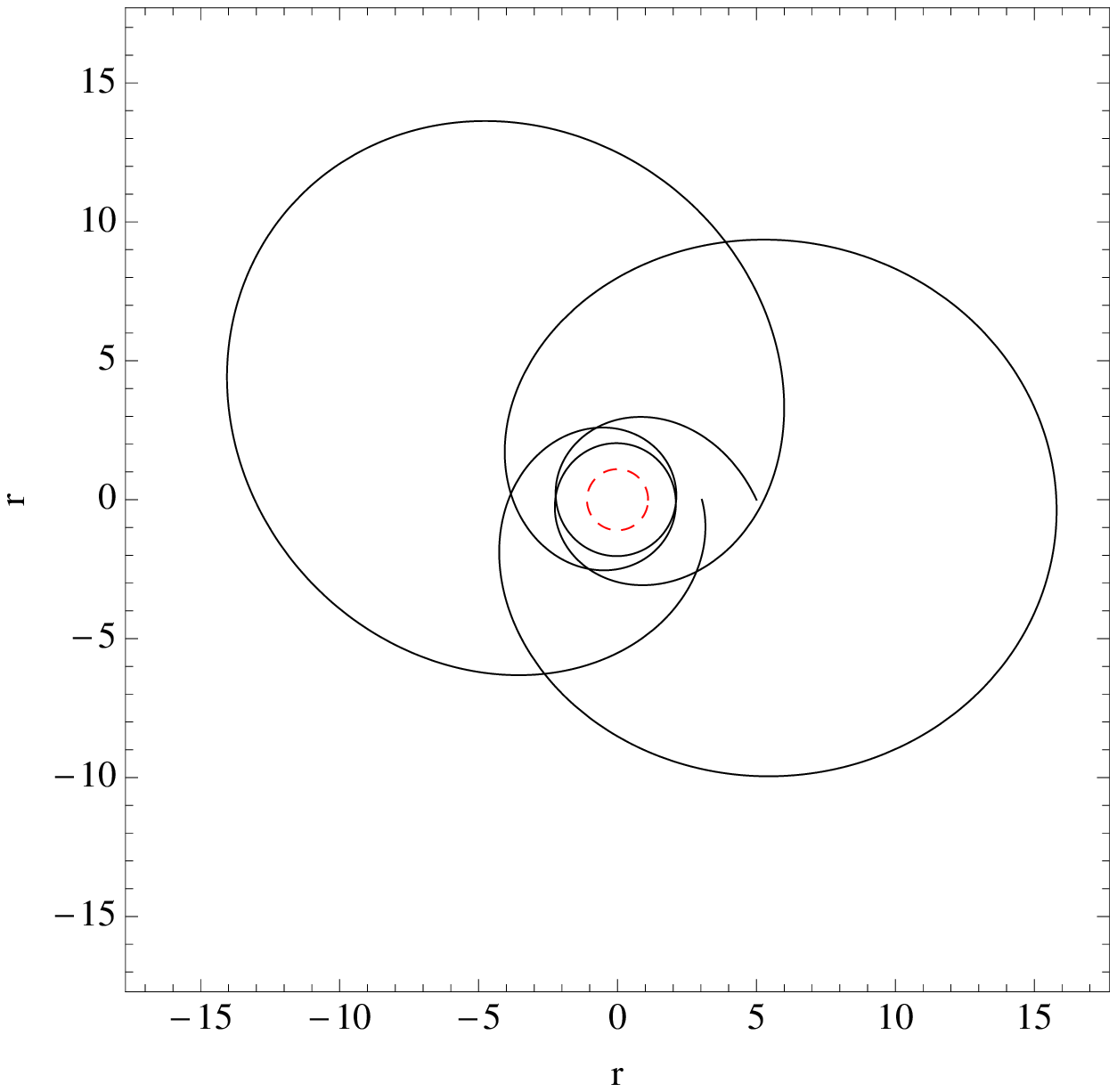}
	\includegraphics[scale=0.406]{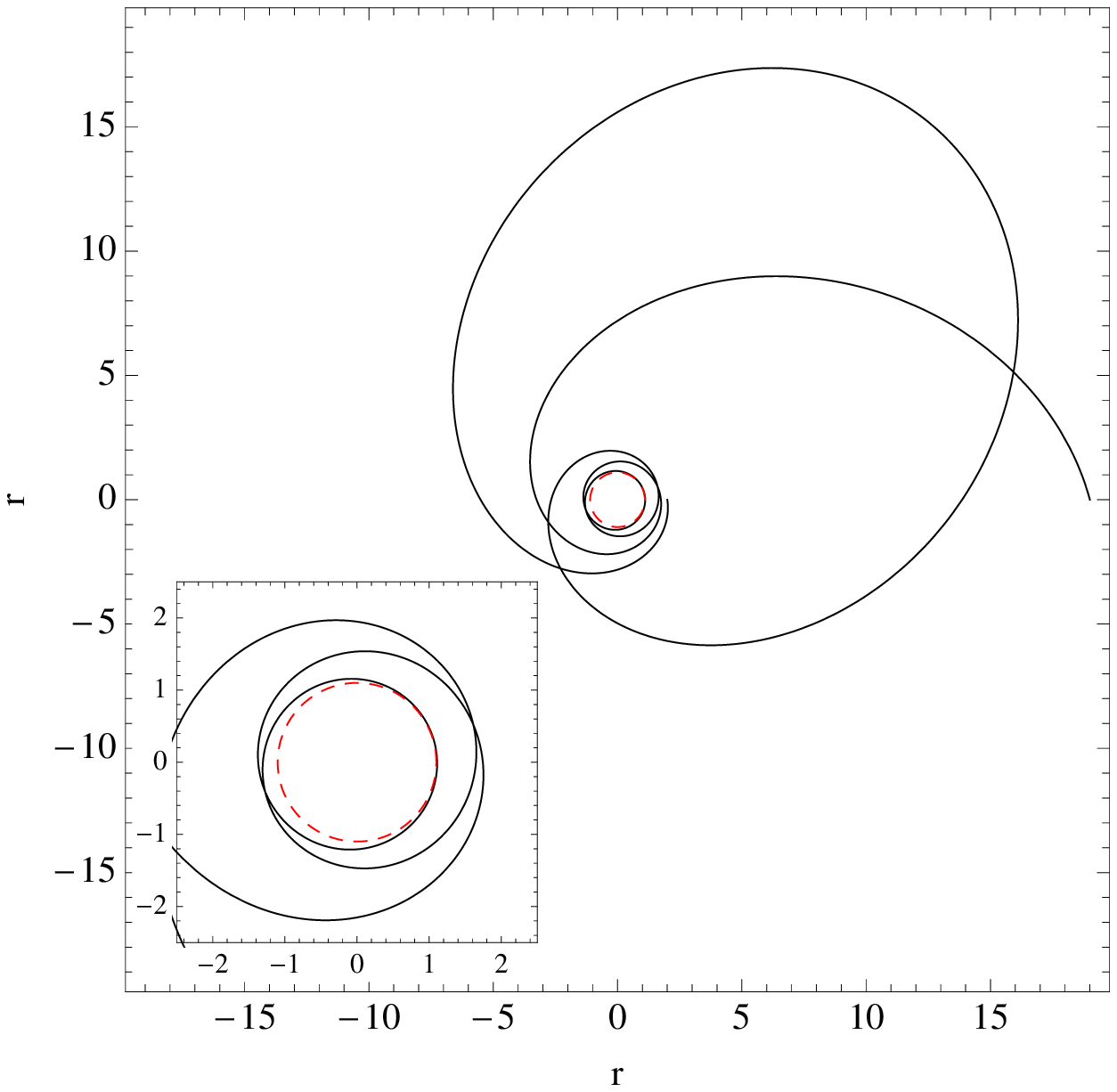}
	\caption{Examples of the time-like bound orbit of the JNW space-time for $E_1^2 =0.92$ and $E_2^2=0.93 $ with fixed $ L=3.6, r_0=2, \mu=2.1$.  }\label{fig:V2 time-like bound orbit}
\end{figure}

Three kinds of bound orbits are plotted in Fig.\ref{fig:V2 time-like bound orbit} with the energy levels between the peak value and bottom values or higher than the peak value of the effective potential. The particle will move on the bound orbit between the range of a perihelion and an aphelion and no particle can fall into the singularity.

\subsubsection{Time-like escape orbit}

\begin{figure}
	\includegraphics[scale=0.4]{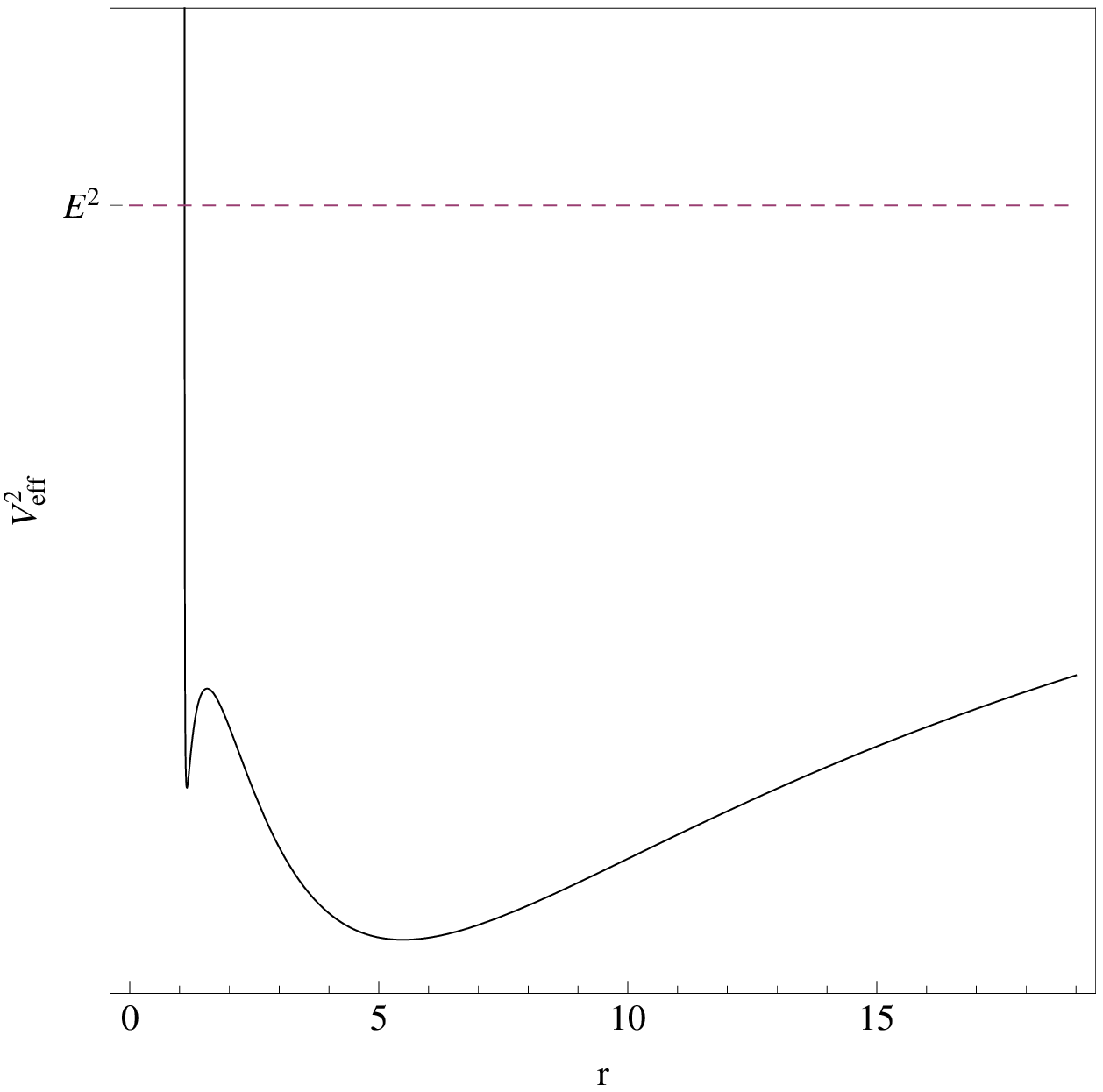}
	\includegraphics[scale=0.41]{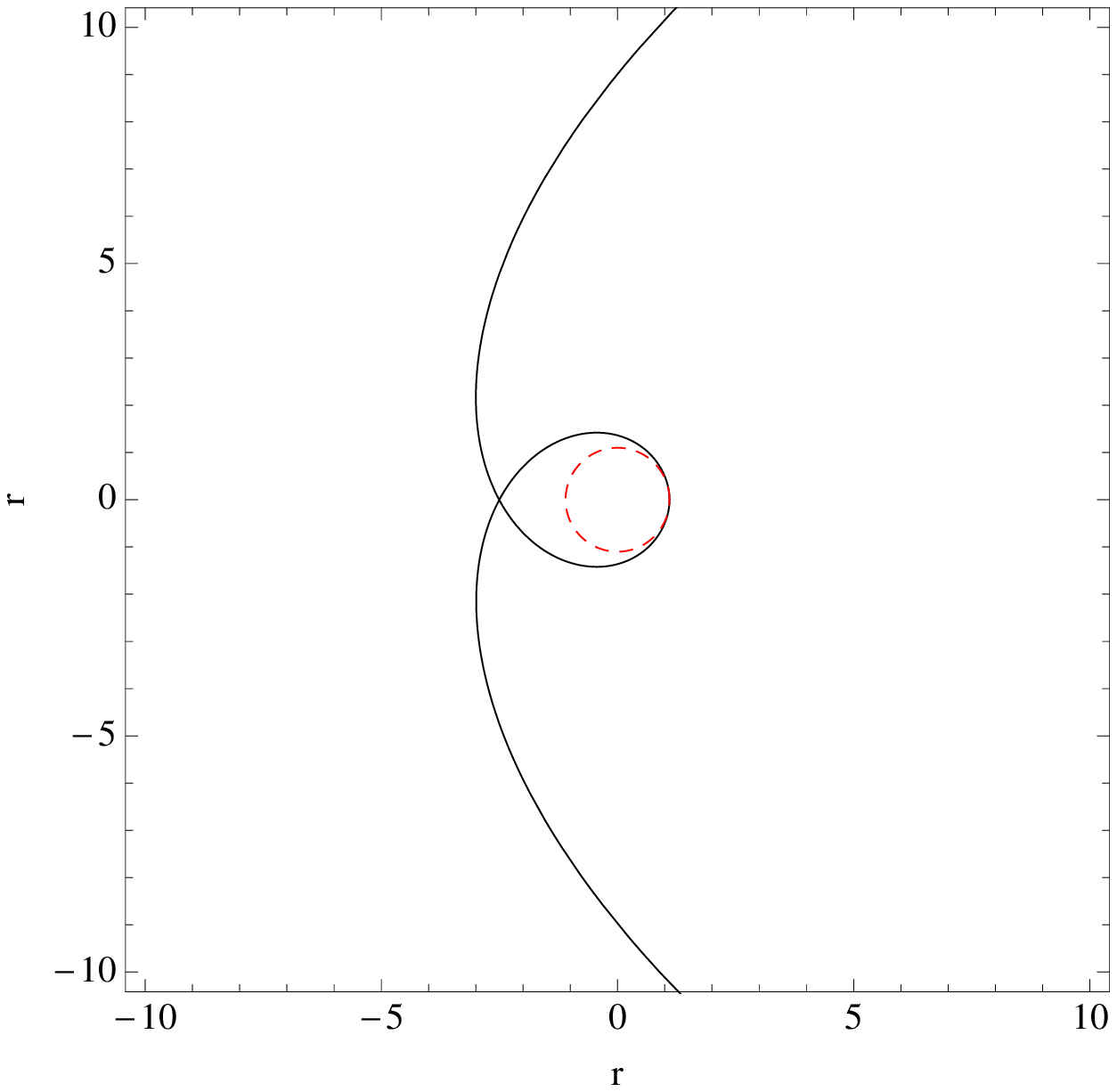}
	\caption{Example of the time-like escape orbit of the JNW space-time with $E^2 =1, L=3.6, r_0=2, \mu=2.1$.  }\label{fig:V2 time-like escape}
\end{figure}

The time-like escape orbit for $\mu=2.1\in(2,\sqrt{5})$ is plotted in Fig.\ref{fig:V2 time-like escape}, the test particle comes from infinity, then reaches a certain distance which is very close to the singularity, at last is reflected back to infinity.

\subsection{Case $\mu \in(\sqrt{5},\infty)$}

\begin{figure}
	\includegraphics[scale=0.6]{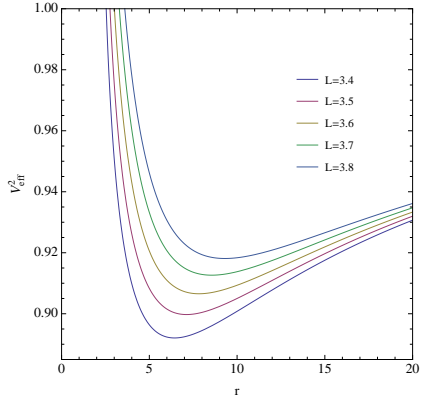}\\
	\caption{The effective potential $V^2_{\rm{eff}}$ of the radial motion is plotted as a function of the radial coordinate $r$ with $\mu =2.82 \in (2,\sqrt{5})$ for different values of $L$. }\label{fig:V-time-like-mu3-L}
\end{figure}

In Fig.\ref{fig:V-time-like-mu3-L} the general behavior of the effective potential is shown as a function of the radius with a fixed value of the parameter  $\mu =2.82 \in (\sqrt{5},\infty)$ for different values of the angular momentum $L$. The figure shows that $V^2_{\rm{eff}}$ has one minimum which indicates the presence of one region of stable circular orbit. And from the effective potential, we can also find the escape orbit of the test particle in the JNW space-time.

\subsubsection{Time-like circular orbit }

\begin{figure}
	\includegraphics[scale=0.4]{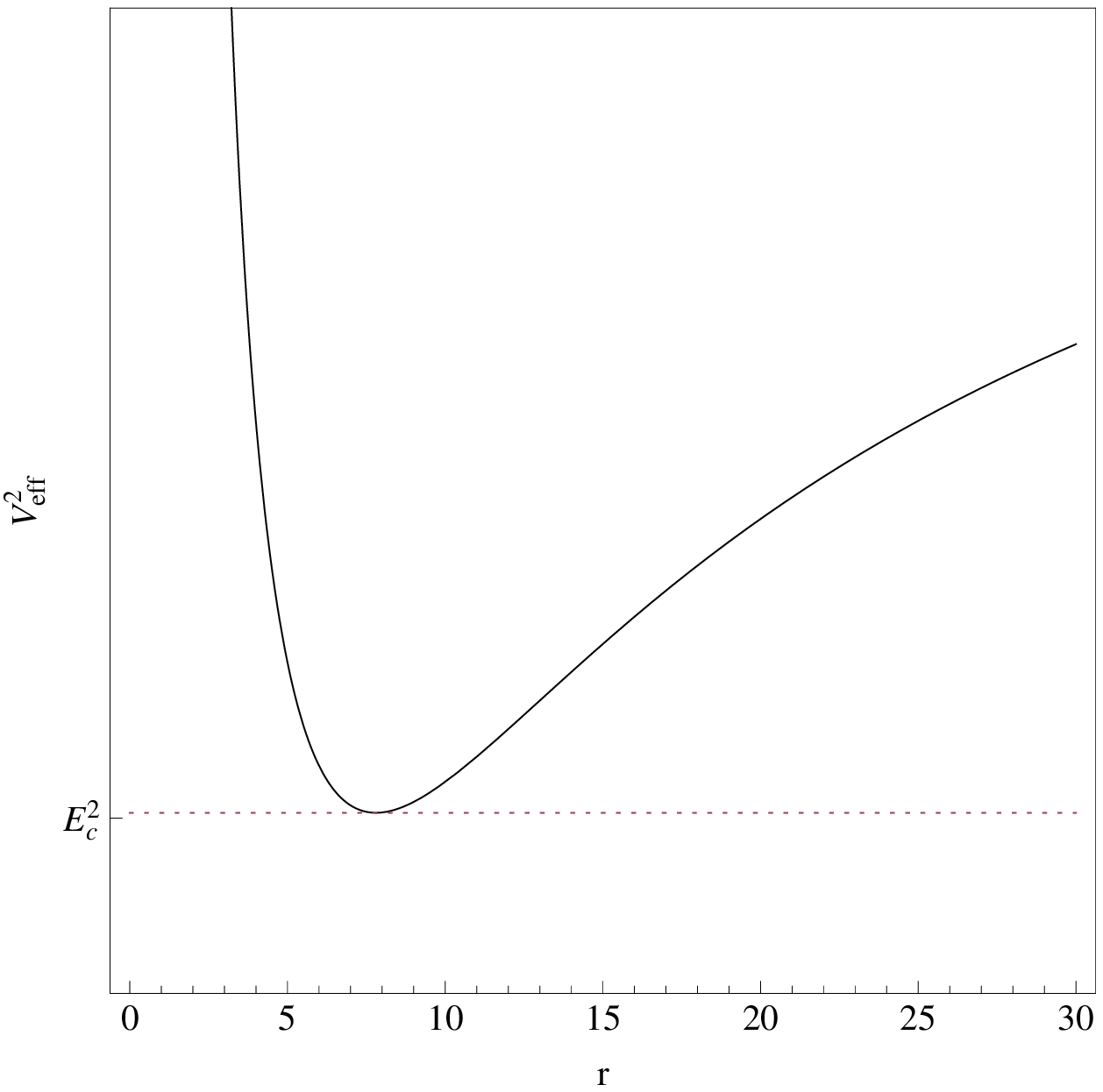}
	\includegraphics[scale=0.4]{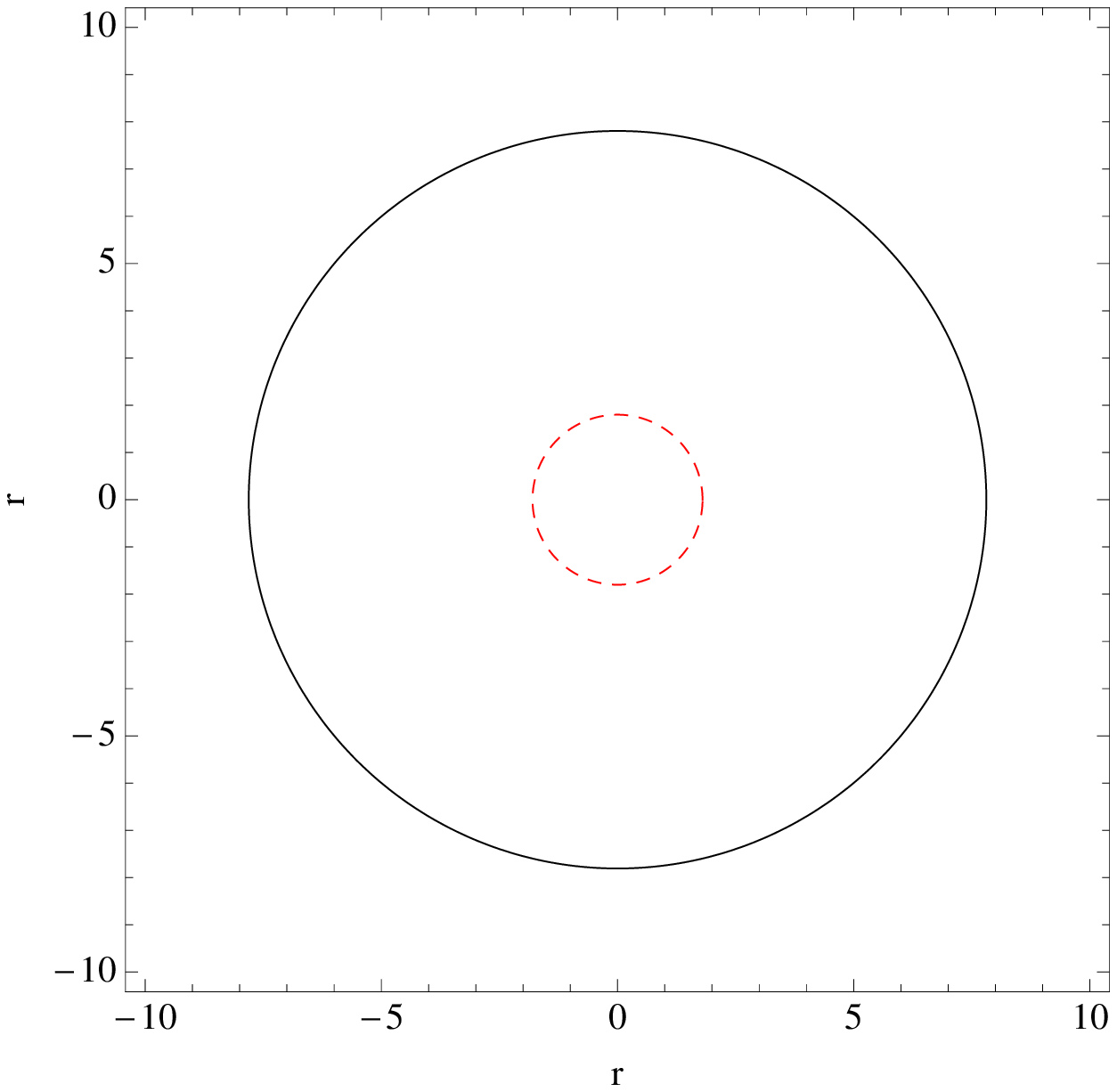}
	\caption{Example of the stable time-like circular geodesics of the JNW space-time with $E_{c}^2 =0.906, L=3.6, r_0=2, \mu=2.8$.  }\label{fig:V3 time-like circle}
\end{figure}

In Fig.\ref{fig:V3 time-like circle} when the energy of the particle is equal to the bottom value of the effective potential curve $E_{c}$, the particle will move on a stable circular orbit. There is  only one circular orbit for this range of the parameter $\mu$.

\subsubsection{Time-like bound orbit}

\begin{figure}
	\includegraphics[scale=0.4]{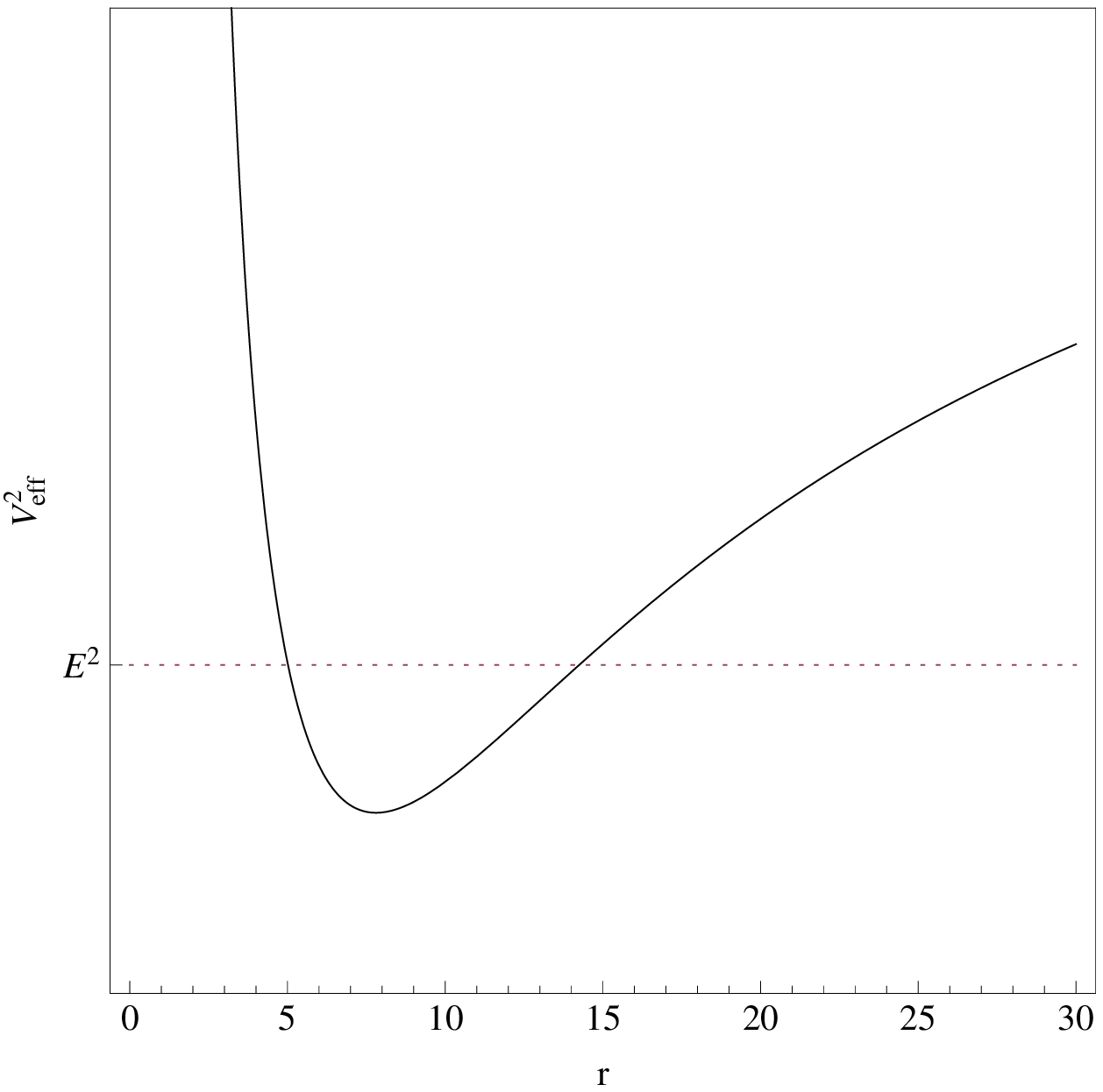}
	\includegraphics[scale=0.4]{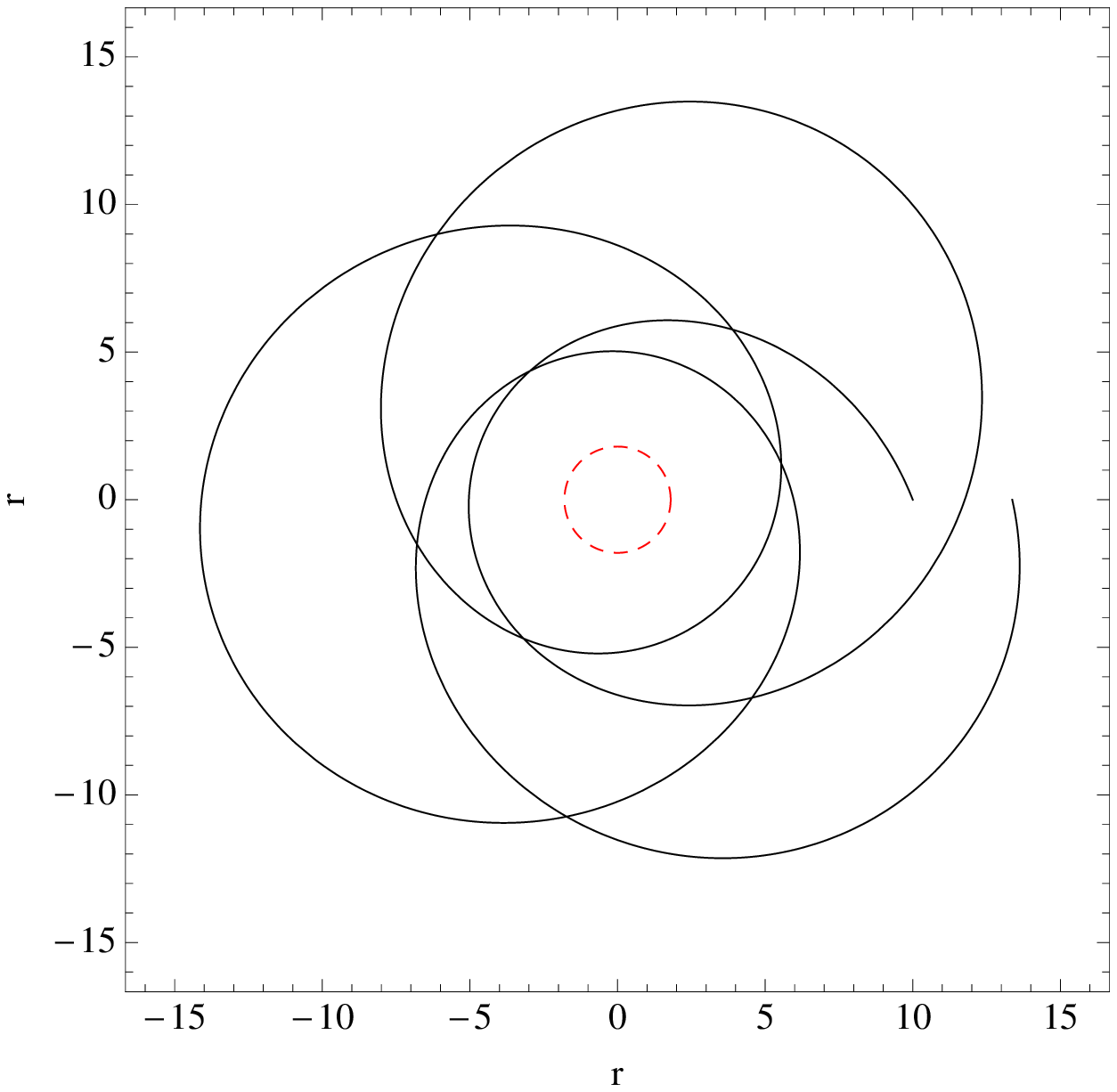}
	\caption{Example of the time-like bound orbit of the JNW space-time with $E^2 =0.92, L=3.6, r_0=2, \mu=2.8$.  }\label{fig:V3 time-like bound}
\end{figure}

In Fig.\ref{fig:V3 time-like bound}, the particle will move on a bound orbit with the radius between an aphelion and a perihelion.

\subsubsection{Escape orbit}

\begin{figure}
	\includegraphics[scale=0.4]{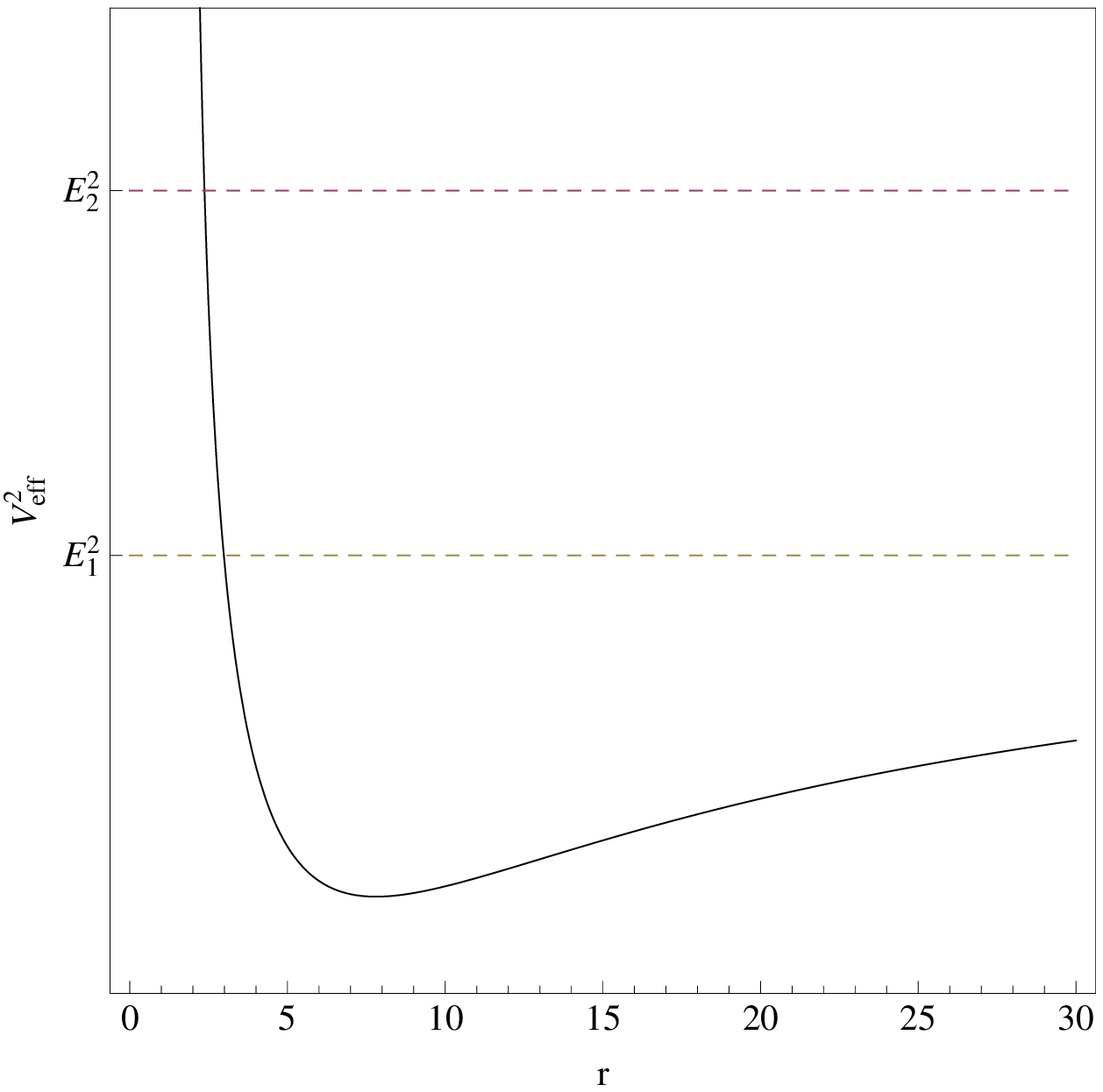}
	\includegraphics[scale=0.4]{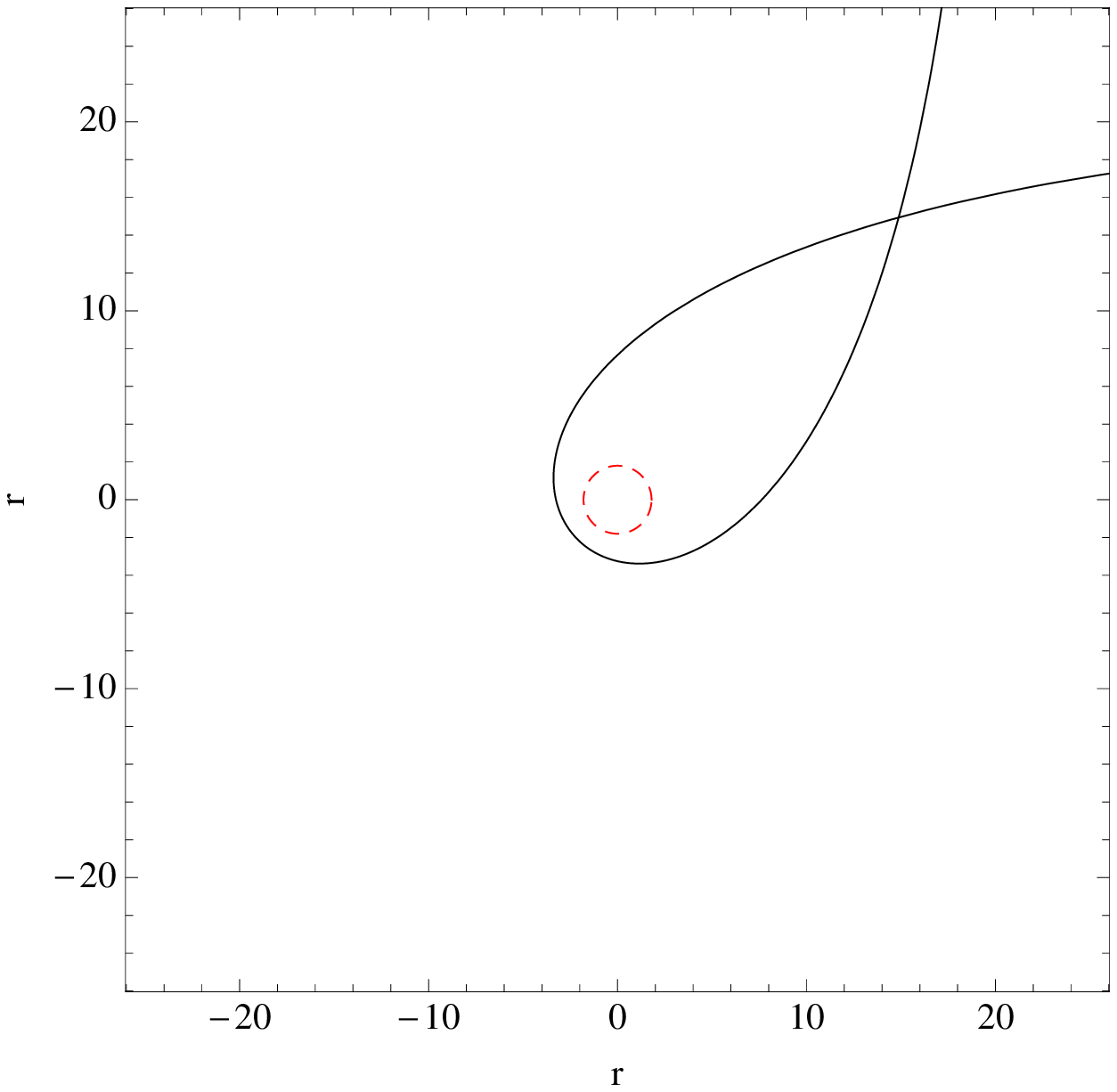}
	\includegraphics[scale=0.4]{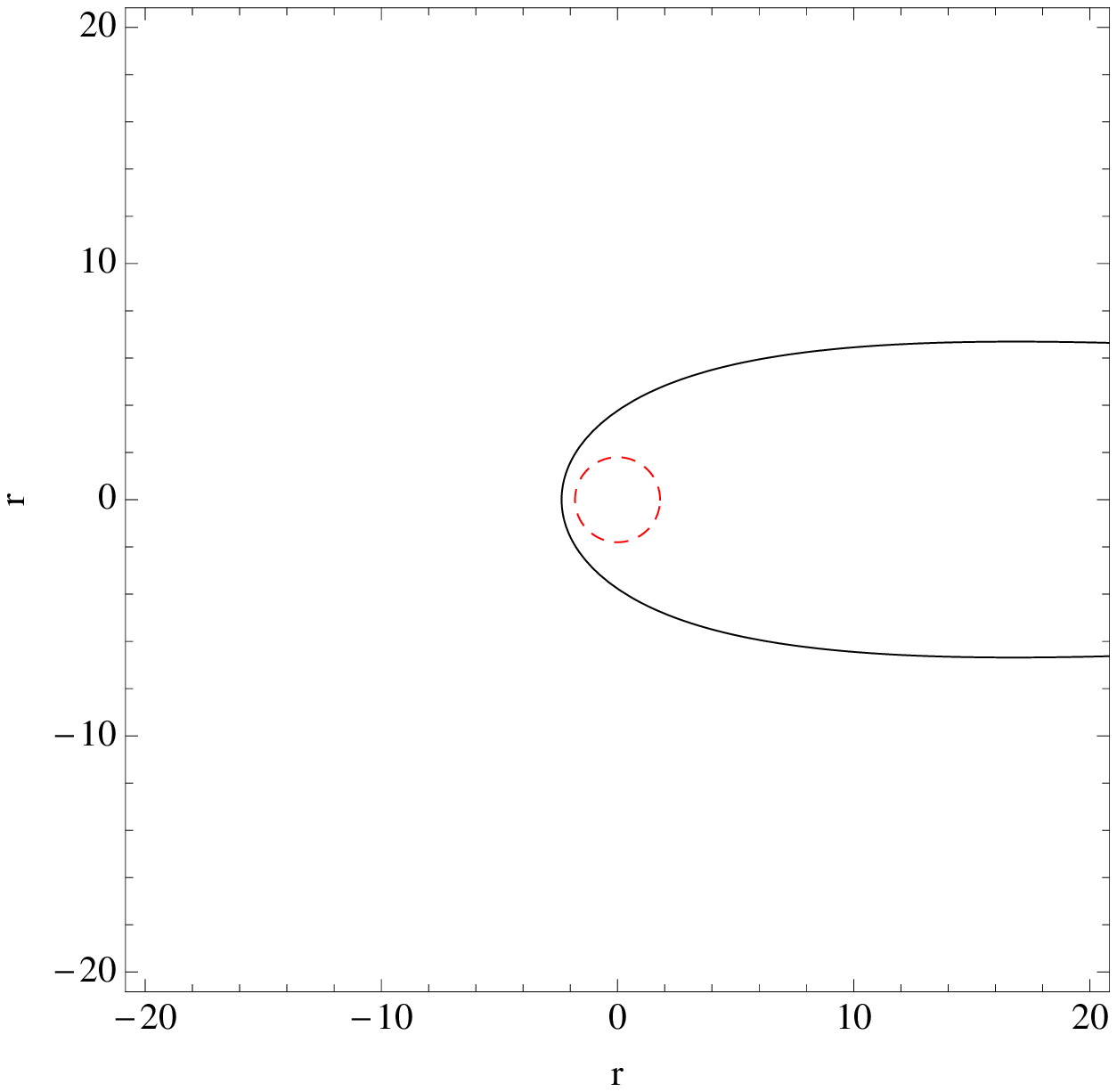}
	\caption{Examples of the time-like escape orbit of the JNW space-time with $E_{1}^2 =1$ and $E_{2}^2 =1.1, L=3.6, r_0=2, \mu=2.8$.  }\label{fig:V3 time-like escape orbit}
\end{figure}

Fig.\ref{fig:V3 time-like escape orbit} shows two escape orbits corresponding to different energy levels. For these both cases, the particle coming from infinity will reach a certain distance closing to the singularity, then escape to infinity, which is reflected by the potential barrier.

\section{NULL GEODESICS}\label{null geodesics}

For the null geodesics $h=0$, the effective potential $V^2_{\rm{eff}}$ becomes

\begin{equation}\label{eq:null effective potential}
V_{\rm{eff}}^2 = A(r)\frac{L^2}{B(r)},
\end{equation}
for circular geodesics
\begin{equation}
\frac{\partial V}{\partial r}=0,
\end{equation}
we have
\begin{equation}\label{eq:circular condition}
4L(r_{0}-2r)A(r)=0,
\end{equation}
where we can see that the null circular geodesics can exist only at the radius $r=\frac{r_{0}}{2}=1$,
which is exactly the radius of the photon sphere, i.e. photon moves on a circular trajectory with a fixed radius, which is independent on the parameter $ \mu $ and energy level.

Imposing the conditions for the stability of the circular geodesics (\ref{stable condition}) in Eq.\ref{eq:null effective potential}, we can get
\begin{equation}\label{inequation for null stability}
8r^2-8rr_0+r^2_0(\mu^2-2)>0
\end{equation}

\begin{figure}
  \includegraphics[scale=0.6]{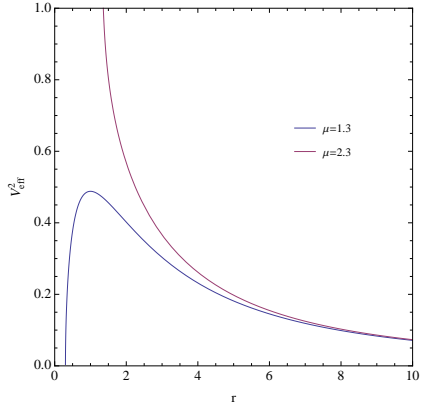}\\
  \caption{The two cases of effective potential $V^2_{\rm{eff}}$ of the radial motion are plotted as a function of the radial coordinate $r$ for $L=3.5, r_0=2,  \mu =1.3 \in (1,2)$ and $\mu =2.3 \in (2,\infty)$, which correspond to two kinds of the different null geodesics. }\label{fig:V-null-mu}
\end{figure}

Solving the above inequality, we find that: I) When $ \mu>2 $, there is always $\frac{\partial^2 V}{\partial r^2}>0$, and  $A(r)>0$, $r>\mu-1>1$, which do not meet the condition for the circular geodesics Eq.(\ref{eq:circular condition}), so there is no null circular geodesics for $\mu $ at the range of $ \mu\in(2,\infty) $; II) When $1<\mu< 2$, there exists an unstable circular orbit at the radius $r=1$. The structure of the null geodesics will be different between two distinct ranges of the parameter $\mu$, i.e. $\mu\in (1,2)$ and $\mu\in (2,\infty)$. Fig.\ref{fig:V-null-mu} shows two kinds of effective potentials of the photon. We will discuss all the possible orbits of the photon corresponding to these two cases.

According to Eqs.(\ref{eq:motion}, \ref{eq:null effective potential}), the motion equation of the photon reads
\begin{equation}\label{null orbit equation}
	\dot{r}^2 = E^2 - A(r)\frac{L^2}{r^2}.
\end{equation}
Using Eq.(\ref{eq:L1}) and making a variable transformation $u^{-1}=r$, Eq. (\ref{null orbit equation}) becomes

\begin{equation}\label{eq:orbit equation for photon}
	(\frac{du}{d\phi})^2=\frac{u^4}{16}(\frac{E^2a^{2-\frac{2}{\mu}}b^{2+\frac{2}{\mu}}}{L^2}-4ab),
\end{equation}
where $a=2u^{-1}+r_0(1-\mu)$ and $b=2u^{-1}+r_0(1+\mu)$.

We solve Eq.(\ref{eq:orbit equation for photon}) numerically to find all types of null geodesics and examine how the
parameter $\mu$ influences on the geodesics in the JNW space-time.

\subsection{Case $\mu \in (1,2)$}

\begin{figure}
	\includegraphics[scale=0.6]{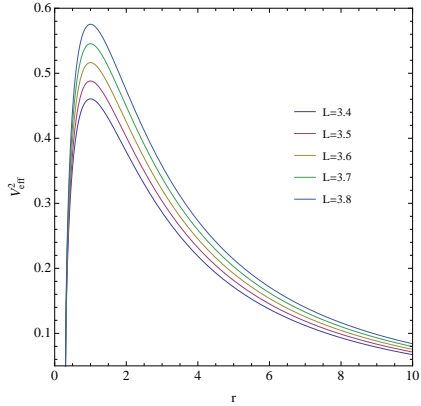}\\
	\caption{The effective potential $V^2_{\rm{eff}}$ of the radial motion is plotted as a function of radial coordinate $r$ with $\mu =1.3 \in (1,2)$ for different values of $L$. It is shown in the figure that $V^2_{\rm{eff}}$ has one maximum at $r=1$, which indicates the presence of one region of unstable circular orbits. }\label{fig:V-mu1-L}
\end{figure}

For this range of the parameter $\mu$, there is only one unstable circular orbit with
fixed radius $r=1$. In Fig.\ref{fig:V-mu1-L} the general behavior of the effective potential is shown as a function of the radius with a fixed value of the parameter  $\mu =1.3 \in (1,2)$ for different values of the angular momentum $L$. The effective potential has one maximum at $r=1$ which corresponds to the unstable circular orbit.

\subsubsection{Null circular orbit}
\begin{figure}
	\includegraphics[scale=0.4]{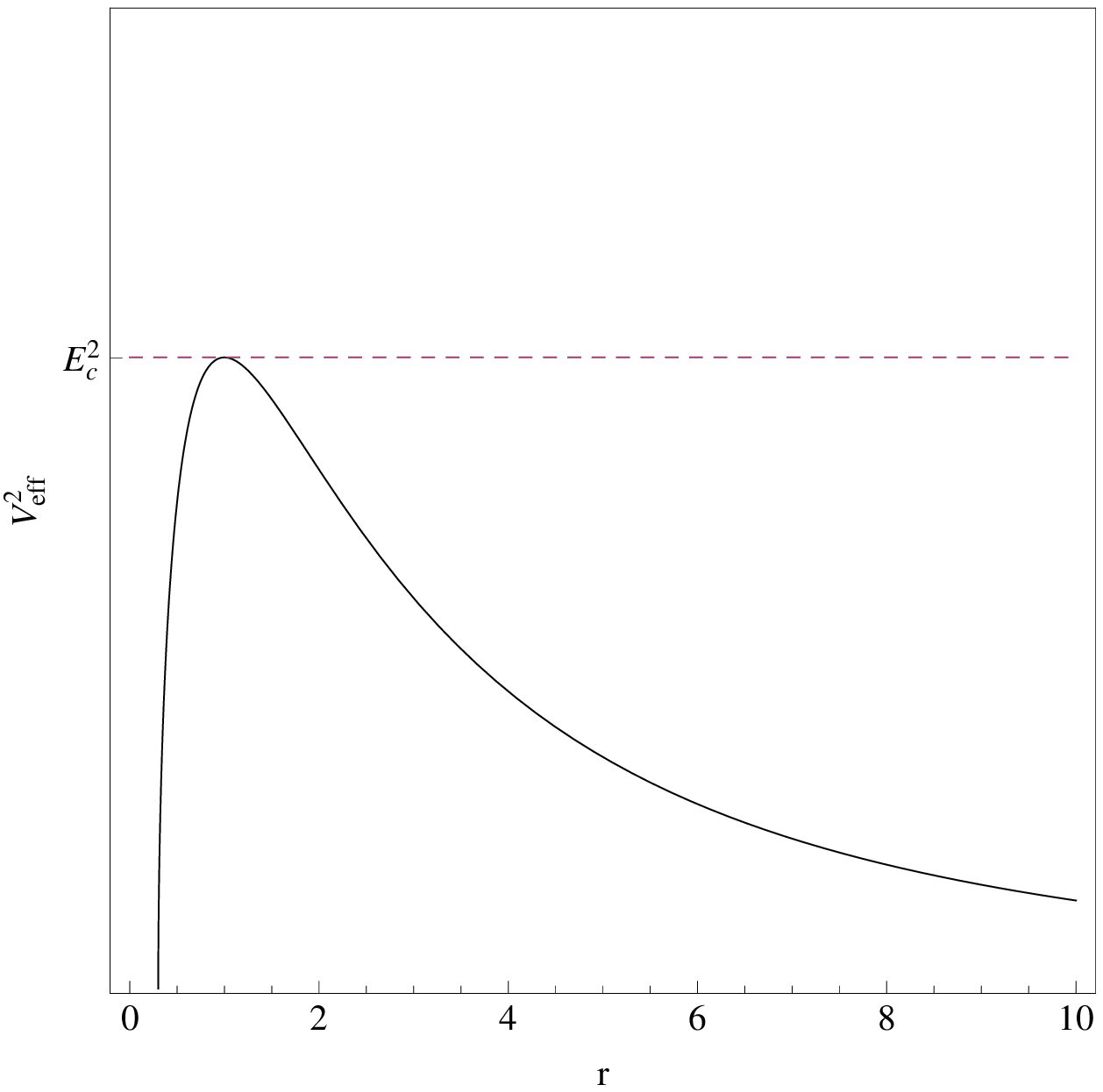}
	\includegraphics[scale=0.4]{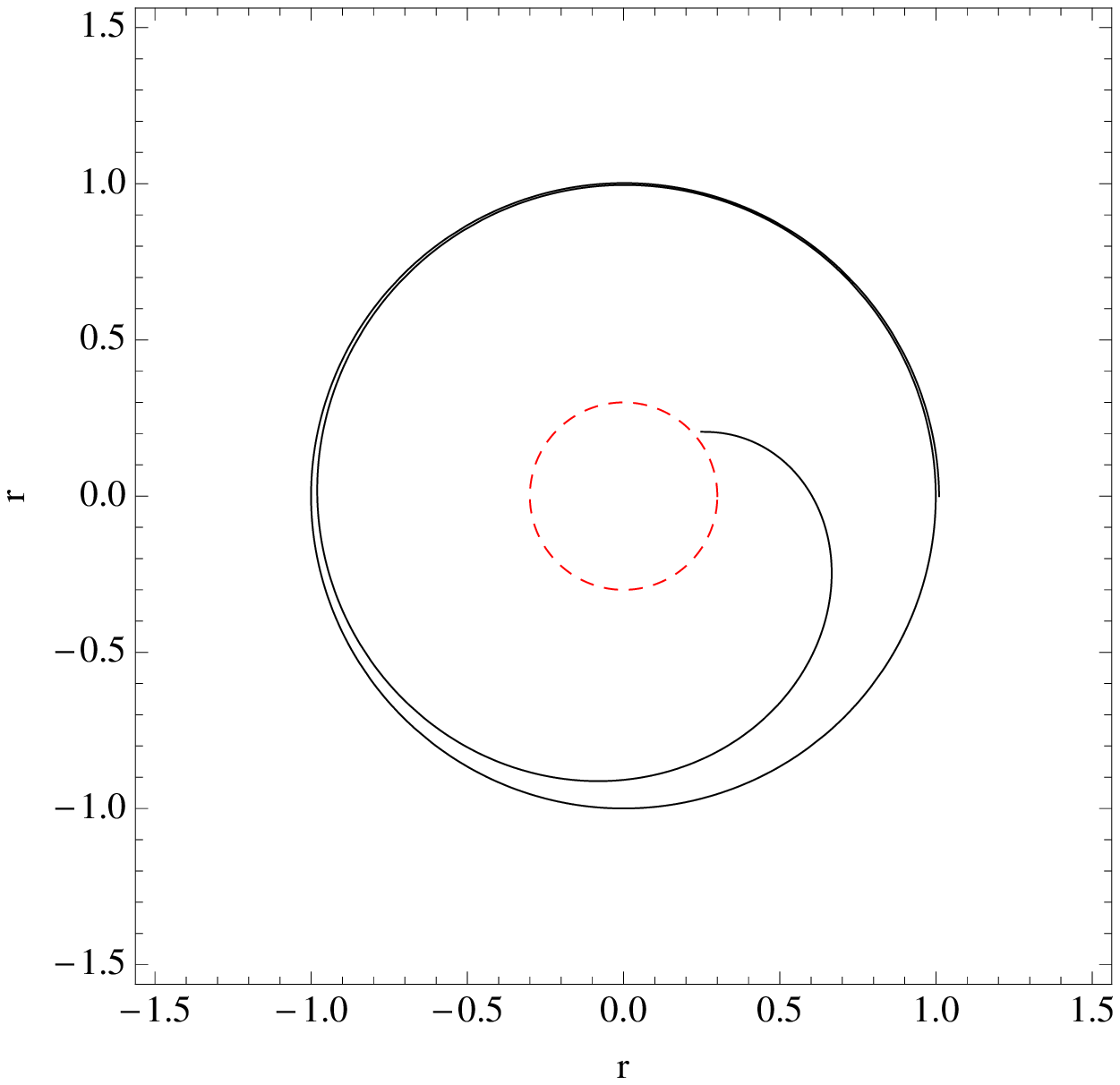}
	\includegraphics[scale=0.4]{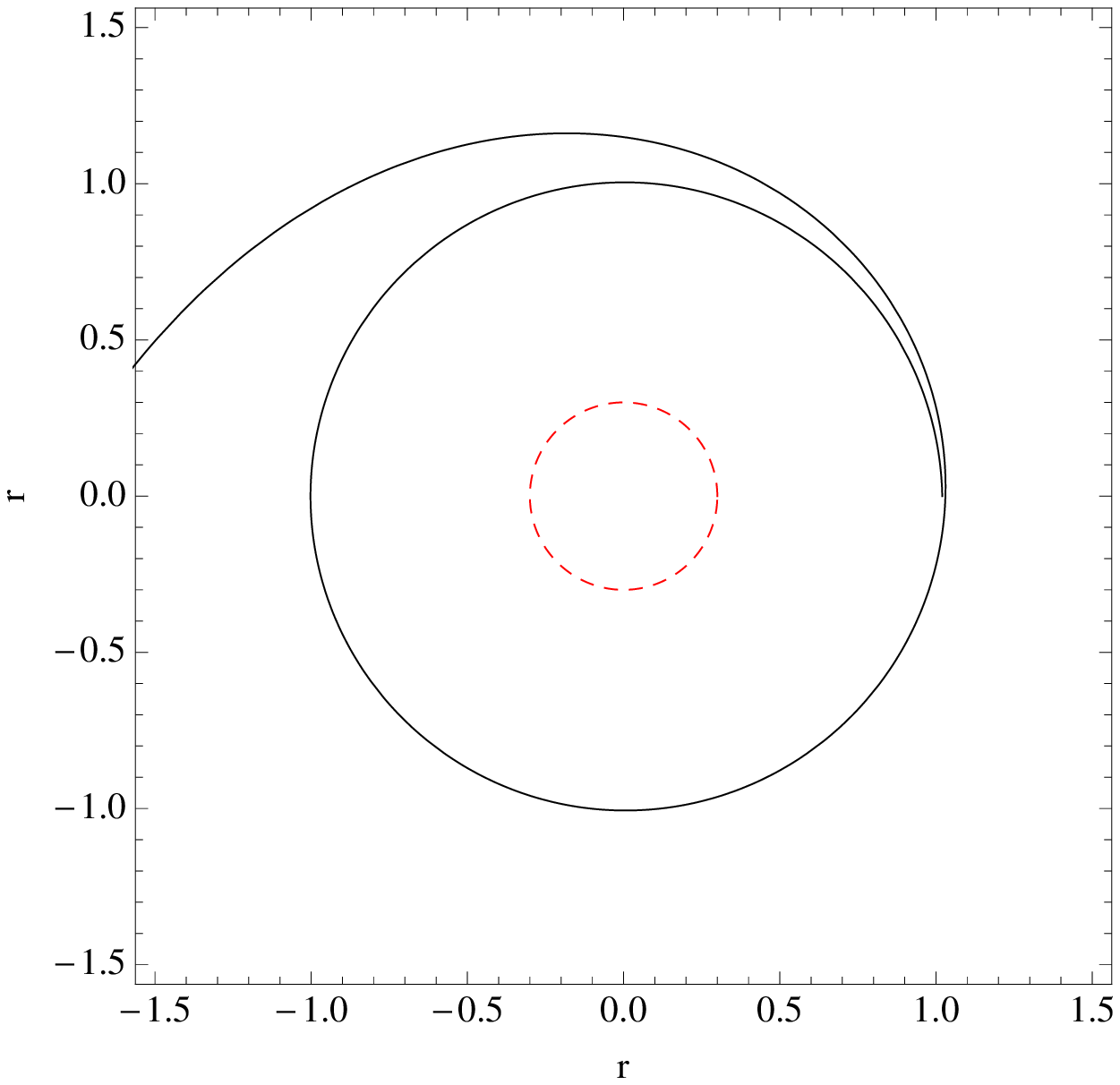}
	\caption{Examples of the unstable null circle geodesics of the JNW space-time with $E^2 =0.516, L=3.6, r_0=2, \mu=1.3$.  }\label{fig:null-circle}
\end{figure}

In Fig.\ref{fig:null-circle}, from the effective potential we can see that there is only one unstable circular orbit at $r=1$. When the energy of the photo is equal to the peak value $E_c$ of the effective potential, the photon is on an unstable circular orbit with radius $r = 1$. Any perturbation will make such unstable orbit recede from the circle to the singularity, which is shown in the middle figure of Figs.\ref{fig:null-circle}, or  escape to infinity on the other side of the potential barrier, which is plotted in the right figure of Fig.\ref{fig:null-circle}.

\subsubsection{Null terminating escape orbit and terminating orbit}

\begin{figure}
	\includegraphics[scale=0.401]{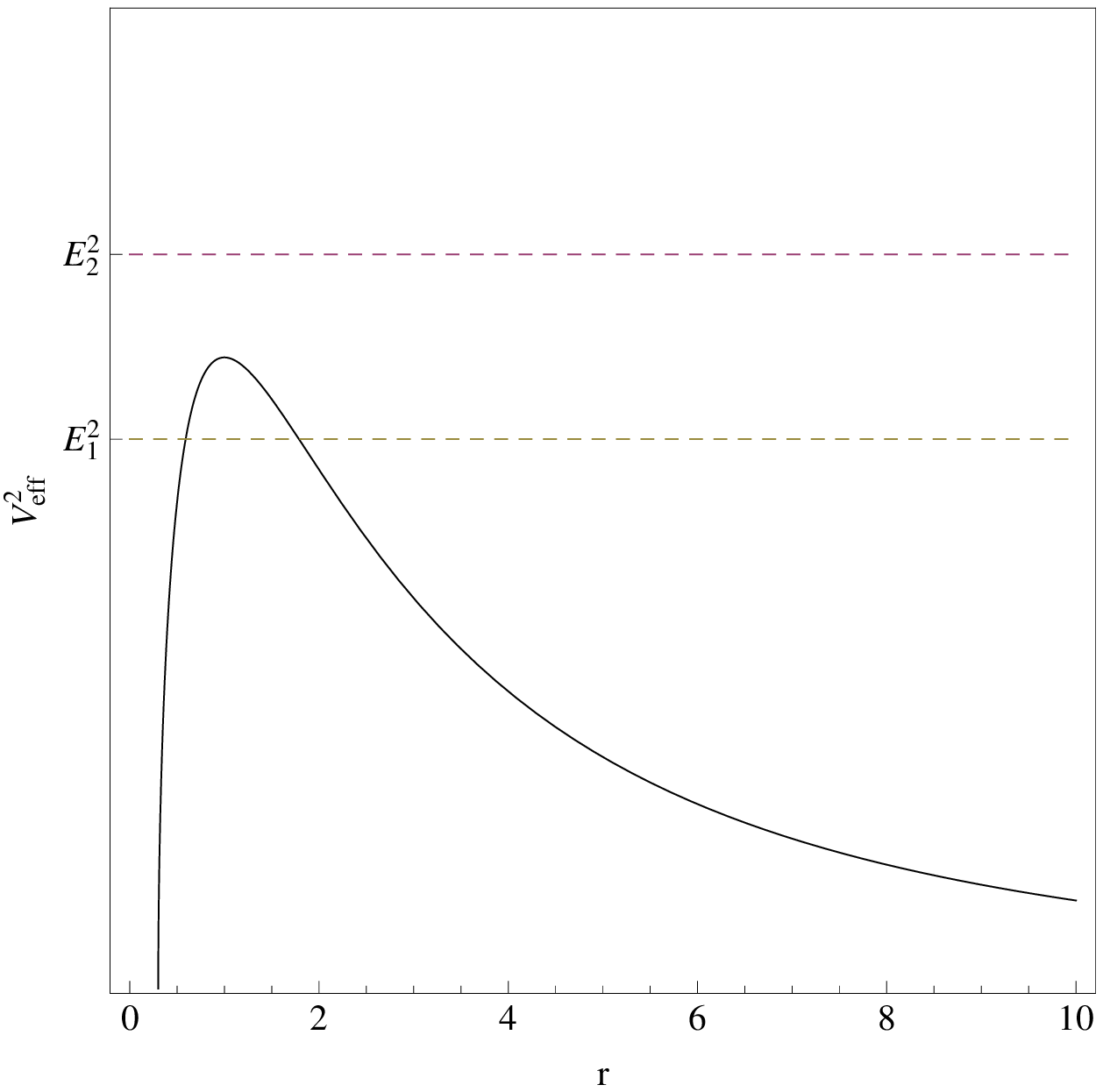}
	\includegraphics[scale=0.4]{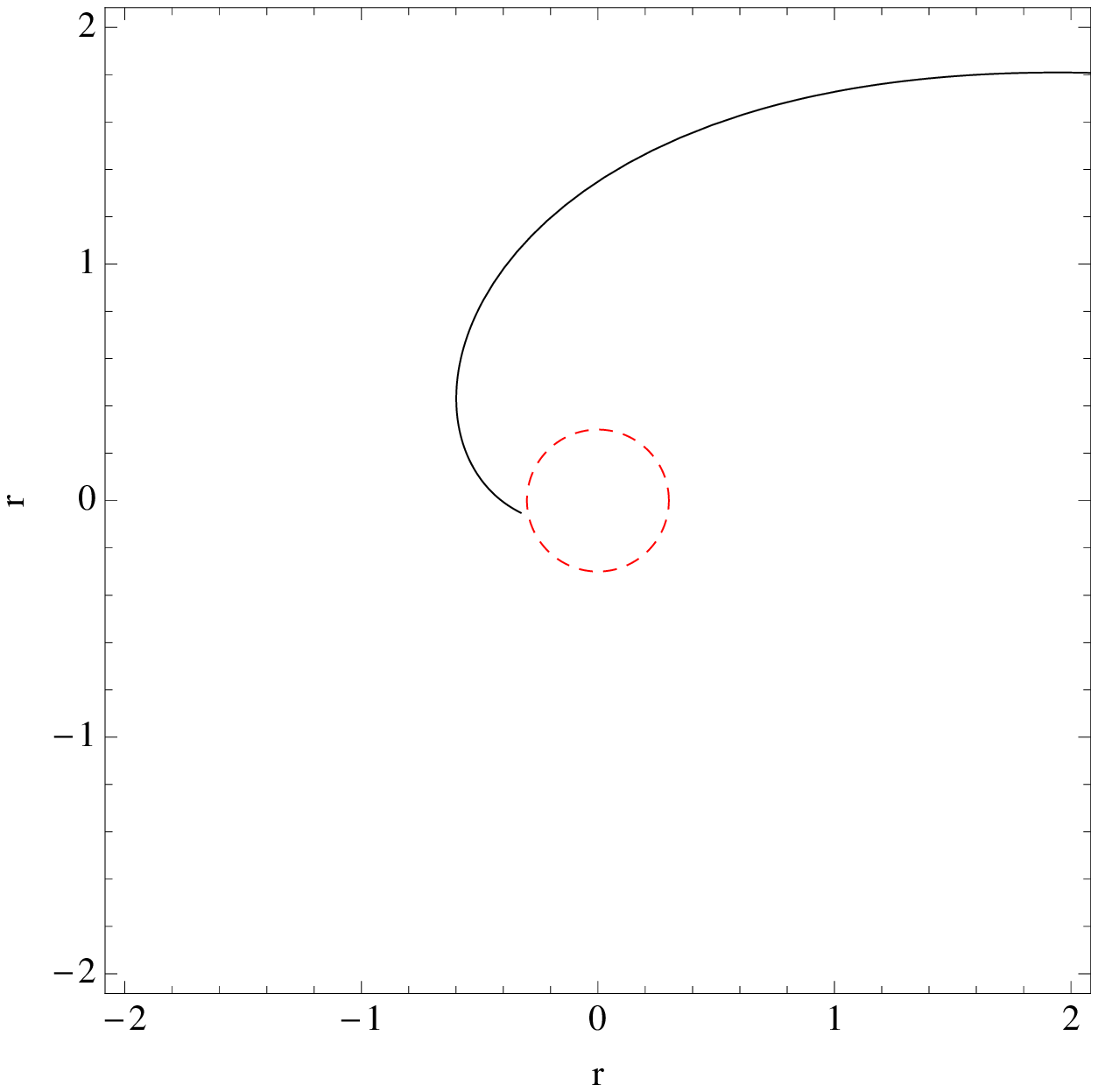}\\
	\includegraphics[scale=0.41]{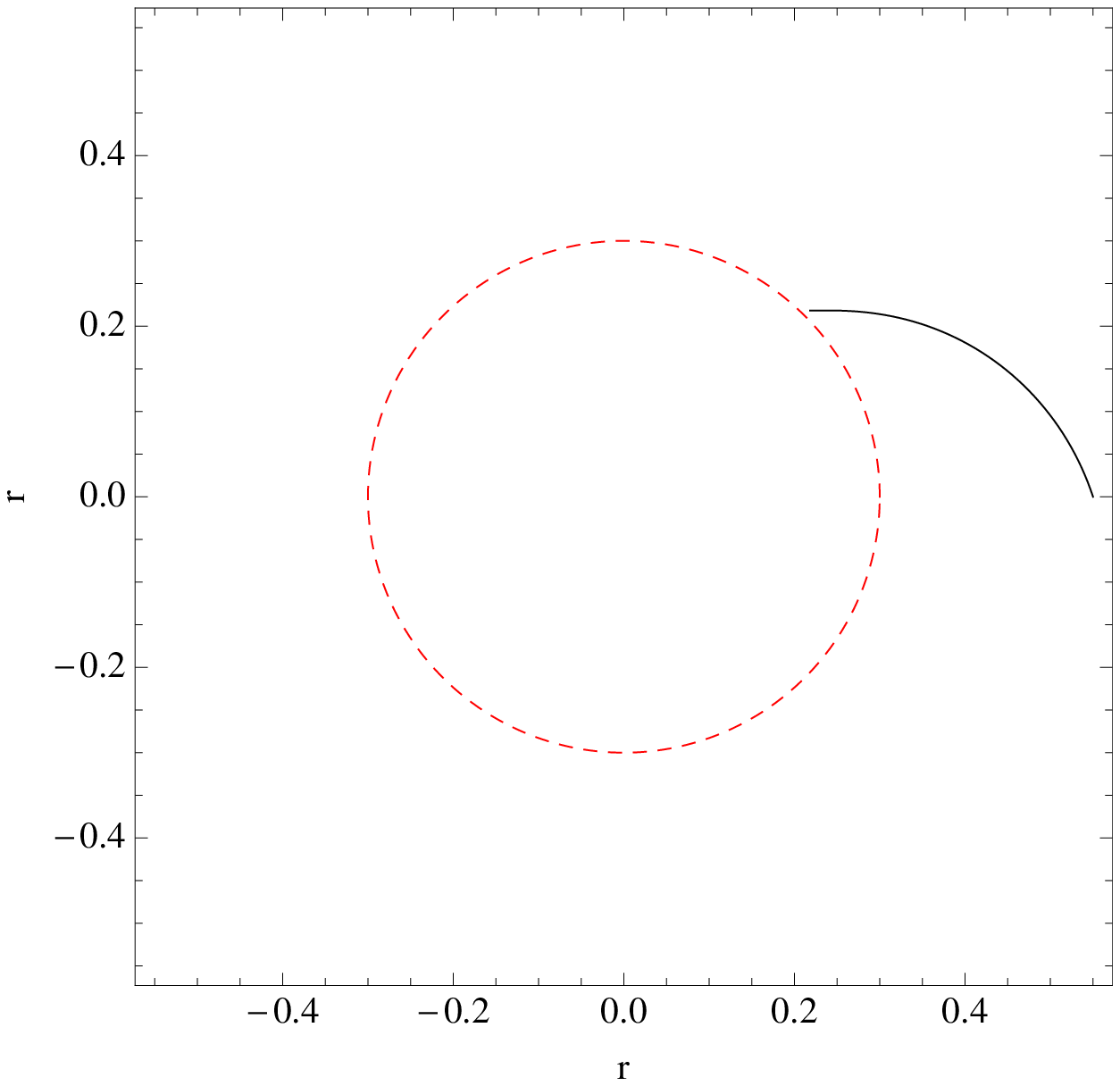}
	\includegraphics[scale=0.4]{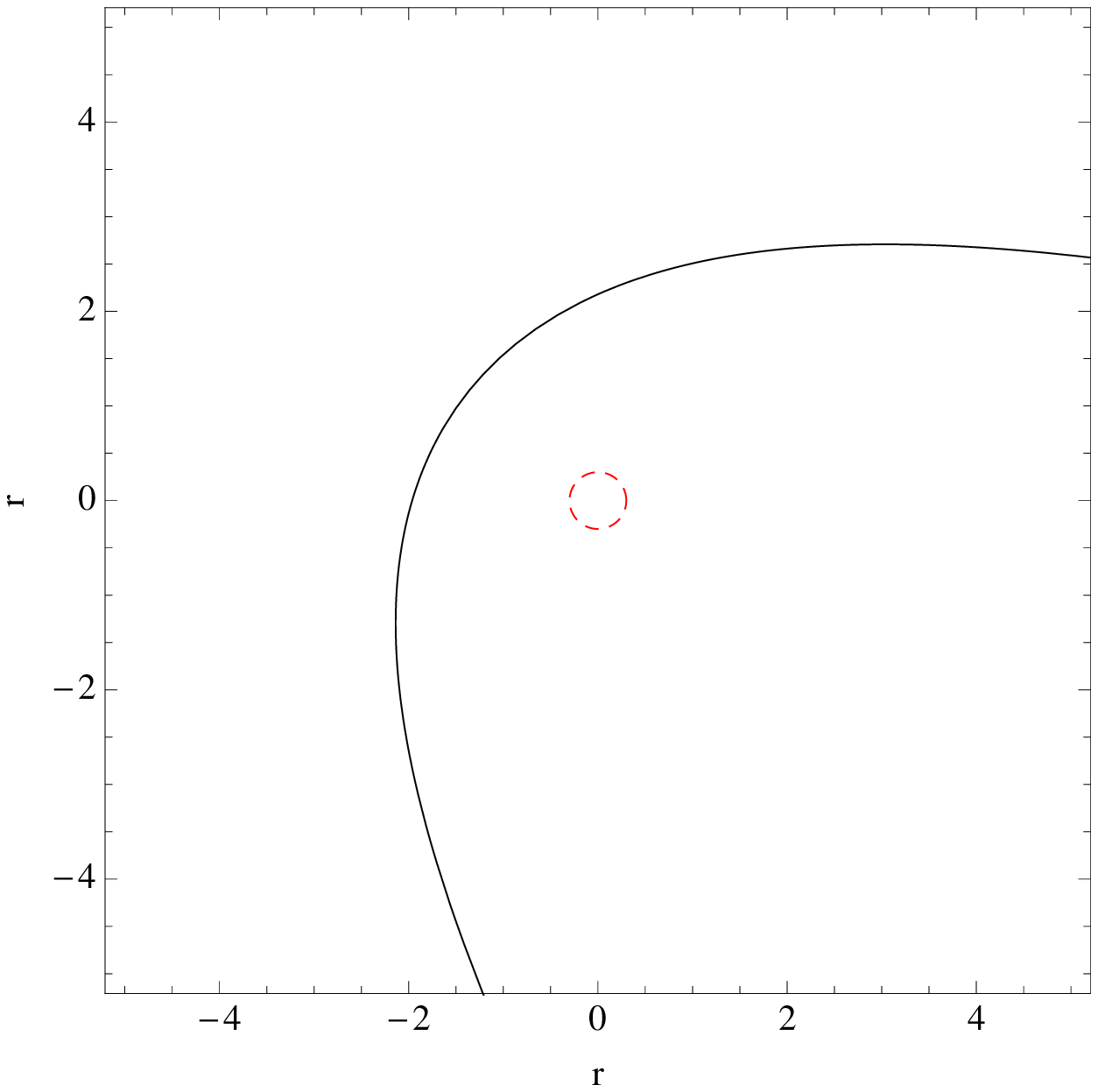}
	\caption{Examples of the null terminating escape orbit, terminating orbit and escape orbit of the JNW space-time with $E_1^2 =0.6, E_2^2 =0.45, L=3.6, r_0=2, \mu=1.3$. }\label{fig:TEO TO}
\end{figure}

A terminating escape orbit (TEO) in the range $r \in (0,\infty)$ exists, whose minimal radius tends to zero, i.e. the photon comes from infinity and ends at the singularity. So the energy must be higher than the peak value of the barrier; A terminating orbit (TO) is an orbit whose minimal radius tends to zero, that's to say, the photon comes from a finite distance and ends at the singularity at $r = 0$. These two kinds of terminating orbits and the corresponding effective potential are plotted in Fig.\ref{fig:TEO TO}, respectively.

\subsubsection{Null escape geodesics}

The escape orbit (EO) is an orbit whose minimal radius is not zero, i.e. the photon comes from infinity or a certain distance away from the singularity and then goes back to infinity. In Fig.\ref{fig:TEO TO}, the energy of the photo $E_1$ is lower than the peak value of the barrier, the photon is just reflected by the potential barrier.

\subsection{Case $\mu \in (2,\infty)$}

\begin{figure}
	\includegraphics[scale=0.6]{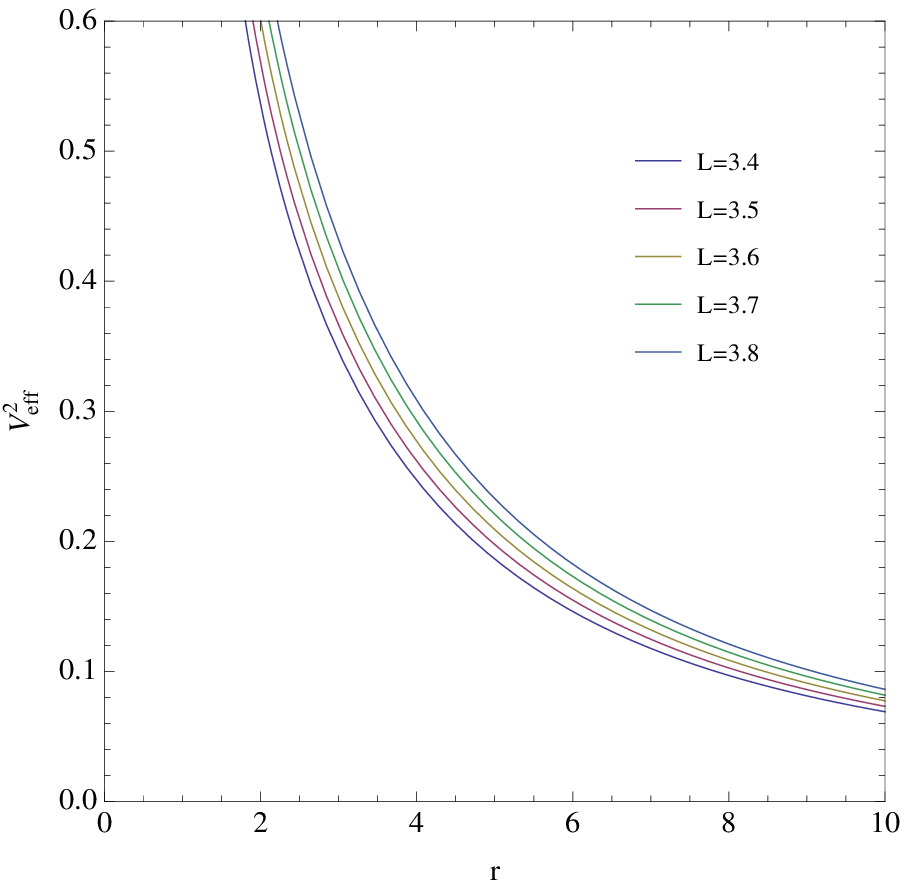}
	\includegraphics[scale=0.4]{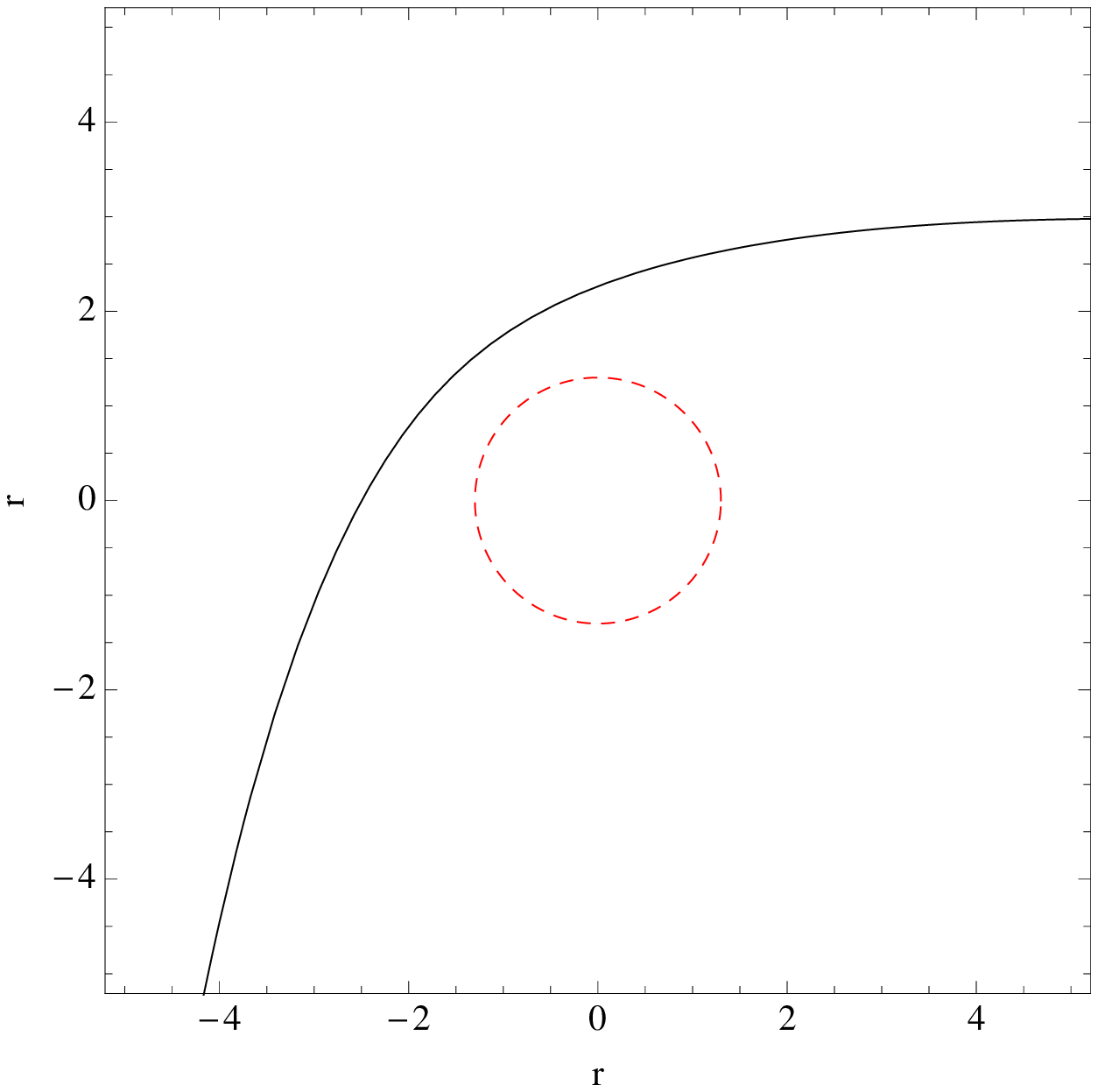}
	\caption{The effective potential $V^2_{\rm{eff}}$ of radial motion is plotted as a function of radial coordinate $r$ for different values of $L$, and the example of the null escape geodesics of the JNW space-time with $E^2 =0.6, L=3.6, r_0=2, \mu=2.3$.}\label{fig:V-null-mu2-L}
\end{figure}

In Fig.\ref{fig:V-null-mu2-L} the  behavior of the effective potential is shown as a function of the radius with a fixed value of the parameter  $\mu =2.3 \in (2,\infty)$ for different values of the angular momentum $L$. We can see that $V^2_{\rm{eff}}$ does not exist neither any maximum nor minimum, which means that neither unstable circular orbit nor stable circular orbit exists for this range of the parameter $\mu$. From the effective potential we can see that there is only escape orbit for this case. The photon, which comes from infinity, reaches a minimum radius and then goes back to infinity. The photon is just reflected by the potential barrier.

\section{Summary and Conclusions}\label{Summary and Conclusions}

In this paper, we have studied the geodesic structure of the JNW space-time which contains a strong curvature naked singularity in detail.  We have solved the geodesic equation and analyzed the behavior of effective potential to investigate the motion of massive and massless particles. By using numerical techniques, we have found that for a test particle I) When $\mu \in (1,2)$, there exist stable and unstable circular orbits, a bound orbit, terminating and terminating escape orbits; II) When $\mu \in (2,\sqrt{5})$, there is an unstable circular orbit between two stable circular orbits, or two kinds of bound orbits, or escape orbits; III) When $\mu \in (\sqrt{5}, \infty)$, the test particle will move on a stable circular orbit, a abound orbit, or an escape orbit. For a photon, I) When $\mu \in (1,2)$, there are unstable circular orbit, terminating and a terminating escape orbit; II) When $\mu \in (2,\infty)$, there is only an escape orbit.

\section{Acknowledgments}

S.Z. would like to thank Jiawei Hu for helpful discussions. This project is supported by the National Natural Science Foundation of China under Grant No.10873004, the State Key Development Program for Basic Research Program of China under Grant No.2010CB832803.

\bibliography{JNWgeodesics}

\begin{thebibliography}{35}%
\makeatletter
\providecommand \@ifxundefined [1]{%
 \@ifx{#1\undefined}
}%
\providecommand \@ifnum [1]{%
 \ifnum #1\expandafter \@firstoftwo
 \else \expandafter \@secondoftwo
 \fi
}%
\providecommand \@ifx [1]{%
 \ifx #1\expandafter \@firstoftwo
 \else \expandafter \@secondoftwo
 \fi
}%
\providecommand \natexlab [1]{#1}%
\providecommand \enquote  [1]{``#1''}%
\providecommand \bibnamefont  [1]{#1}%
\providecommand \bibfnamefont [1]{#1}%
\providecommand \citenamefont [1]{#1}%
\providecommand \href@noop [0]{\@secondoftwo}%
\providecommand \href [0]{\begingroup \@sanitize@url \@href}%
\providecommand \@href[1]{\@@startlink{#1}\@@href}%
\providecommand \@@href[1]{\endgroup#1\@@endlink}%
\providecommand \@sanitize@url [0]{\catcode `\\12\catcode `\$12\catcode
  `\&12\catcode `\#12\catcode `\^12\catcode `\_12\catcode `\%12\relax}%
\providecommand \@@startlink[1]{}%
\providecommand \@@endlink[0]{}%
\providecommand \url  [0]{\begingroup\@sanitize@url \@url }%
\providecommand \@url [1]{\endgroup\@href {#1}{\urlprefix }}%
\providecommand \urlprefix  [0]{URL }%
\providecommand \Eprint [0]{\href }%
\providecommand \doibase [0]{http://dx.doi.org/}%
\providecommand \selectlanguage [0]{\@gobble}%
\providecommand \bibinfo  [0]{\@secondoftwo}%
\providecommand \bibfield  [0]{\@secondoftwo}%
\providecommand \translation [1]{[#1]}%
\providecommand \BibitemOpen [0]{}%
\providecommand \bibitemStop [0]{}%
\providecommand \bibitemNoStop [0]{.\EOS\space}%
\providecommand \EOS [0]{\spacefactor3000\relax}%
\providecommand \BibitemShut  [1]{\csname bibitem#1\endcsname}%
\let\auto@bib@innerbib\@empty
\bibitem [{\citenamefont {Podolsky}(1999)}]{Pod99}%
  \BibitemOpen
  \bibfield  {author} {\bibinfo {author} {\bibfnamefont {J.}~\bibnamefont
  {Podolsky}},\ }\href {\doibase 10.1023/A:1026762116655} {\bibfield  {journal}
  {\bibinfo  {journal} {Gen. Rel. Grav.}\ }\textbf {\bibinfo {volume} {31}},\
  \bibinfo {pages} {1703} (\bibinfo {year} {1999})}\BibitemShut {NoStop}%
\bibitem [{\citenamefont {Cruz}\ \emph {et~al.}(2005)\citenamefont {Cruz},
  \citenamefont {Olivares},\ and\ \citenamefont
  {Villanueva}}]{0264-9381-22-6-016}%
  \BibitemOpen
  \bibfield  {author} {\bibinfo {author} {\bibfnamefont {N.}~\bibnamefont
  {Cruz}}, \bibinfo {author} {\bibfnamefont {M.}~\bibnamefont {Olivares}}, \
  and\ \bibinfo {author} {\bibfnamefont {J.~R.}\ \bibnamefont {Villanueva}},\
  }\href {http://stacks.iop.org/0264-9381/22/i=6/a=016} {\bibfield  {journal}
  {\bibinfo  {journal} {Class. Quant. Grav.}\ }\textbf {\bibinfo {volume}
  {22}},\ \bibinfo {pages} {1167} (\bibinfo {year} {2005})}\BibitemShut
  {NoStop}%
\bibitem [{\citenamefont {Pradhan}\ and\ \citenamefont
  {Majumdar}(2011)}]{Pradhan2011474}%
  \BibitemOpen
  \bibfield  {author} {\bibinfo {author} {\bibfnamefont {P.}~\bibnamefont
  {Pradhan}}\ and\ \bibinfo {author} {\bibfnamefont {P.}~\bibnamefont
  {Majumdar}},\ }\href {\doibase
  http://dx.doi.org/10.1016/j.physleta.2010.11.015} {\bibfield  {journal}
  {\bibinfo  {journal} {Phys. Lett. A}\ }\textbf {\bibinfo {volume} {375}},\
  \bibinfo {pages} {474 } (\bibinfo {year} {2011})}\BibitemShut {NoStop}%
\bibitem [{\citenamefont {Pugliese}\ \emph
  {et~al.}(2011{\natexlab{a}})\citenamefont {Pugliese}, \citenamefont
  {Quevedo},\ and\ \citenamefont {Ruffini}}]{PhysRevD.83.104052}%
  \BibitemOpen
  \bibfield  {author} {\bibinfo {author} {\bibfnamefont {D.}~\bibnamefont
  {Pugliese}}, \bibinfo {author} {\bibfnamefont {H.}~\bibnamefont {Quevedo}}, \
  and\ \bibinfo {author} {\bibfnamefont {R.}~\bibnamefont {Ruffini}},\ }\href
  {\doibase 10.1103/PhysRevD.83.104052} {\bibfield  {journal} {\bibinfo
  {journal} {Phys. Rev. D}\ }\textbf {\bibinfo {volume} {83}},\ \bibinfo
  {pages} {104052} (\bibinfo {year} {2011}{\natexlab{a}})}\BibitemShut
  {NoStop}%
\bibitem [{\citenamefont {Pugliese}\ \emph {et~al.}(2013)\citenamefont
  {Pugliese}, \citenamefont {Quevedo},\ and\ \citenamefont
  {Ruffini}}]{reissner-circular}%
  \BibitemOpen
  \bibfield  {author} {\bibinfo {author} {\bibfnamefont {D.}~\bibnamefont
  {Pugliese}}, \bibinfo {author} {\bibfnamefont {H.}~\bibnamefont {Quevedo}}, \
  and\ \bibinfo {author} {\bibfnamefont {R.}~\bibnamefont {Ruffini}},\ }\href
  {http://arXiv.org/abs/1304.2940} {\bibfield  {journal} {\bibinfo  {journal}
  {arXiv:1304.2940}\ } (\bibinfo {year} {2013})}\BibitemShut {NoStop}%
\bibitem [{\citenamefont {Pugliese}\ \emph
  {et~al.}(2011{\natexlab{b}})\citenamefont {Pugliese}, \citenamefont
  {Quevedo},\ and\ \citenamefont {Ruffini}}]{PhysRevD.84.044030}%
  \BibitemOpen
  \bibfield  {author} {\bibinfo {author} {\bibfnamefont {D.}~\bibnamefont
  {Pugliese}}, \bibinfo {author} {\bibfnamefont {H.}~\bibnamefont {Quevedo}}, \
  and\ \bibinfo {author} {\bibfnamefont {R.}~\bibnamefont {Ruffini}},\ }\href
  {\doibase 10.1103/PhysRevD.84.044030} {\bibfield  {journal} {\bibinfo
  {journal} {Phys. Rev. D}\ }\textbf {\bibinfo {volume} {84}},\ \bibinfo
  {pages} {044030} (\bibinfo {year} {2011}{\natexlab{b}})}\BibitemShut
  {NoStop}%
\bibitem [{\citenamefont {Zhou}\ \emph {et~al.}(2011)\citenamefont {Zhou},
  \citenamefont {Chen},\ and\ \citenamefont {Wang}}]{geoBraneShengZ}%
  \BibitemOpen
  \bibfield  {author} {\bibinfo {author} {\bibfnamefont {S.}~\bibnamefont
  {Zhou}}, \bibinfo {author} {\bibfnamefont {J.~H.}\ \bibnamefont {Chen}}, \
  and\ \bibinfo {author} {\bibfnamefont {Y.~J.}\ \bibnamefont {Wang}},\ }\href
  {http://stacks.iop.org/1674-1056/20/i=10/a=100401} {\bibfield  {journal}
  {\bibinfo  {journal} {Chin. Phys. B}\ }\textbf {\bibinfo {volume} {20}},\
  \bibinfo {pages} {100401} (\bibinfo {year} {2011})}\BibitemShut {NoStop}%
\bibitem [{\citenamefont {Zhou}\ \emph {et~al.}(2012)\citenamefont {Zhou},
  \citenamefont {Chen},\ and\ \citenamefont {Wang}}]{geodesicBardeen}%
  \BibitemOpen
  \bibfield  {author} {\bibinfo {author} {\bibfnamefont {S.}~\bibnamefont
  {Zhou}}, \bibinfo {author} {\bibfnamefont {J.~H.}\ \bibnamefont {Chen}}, \
  and\ \bibinfo {author} {\bibfnamefont {Y.~J.}\ \bibnamefont {Wang}},\ }\href
  {\doibase 10.1142/S0218271812500770} {\bibfield  {journal} {\bibinfo
  {journal} {Int. J. Mod. Phys. D}\ }\textbf {\bibinfo {volume} {21}},\
  \bibinfo {pages} {1250077} (\bibinfo {year} {2012})}\BibitemShut {NoStop}%
\bibitem [{\citenamefont {Joshi}\ \emph {et~al.}(2002)\citenamefont {Joshi},
  \citenamefont {Dadhich},\ and\ \citenamefont
  {Maartens}}]{PhysRevD.65.101501}%
  \BibitemOpen
  \bibfield  {author} {\bibinfo {author} {\bibfnamefont {P.~S.}\ \bibnamefont
  {Joshi}}, \bibinfo {author} {\bibfnamefont {N.}~\bibnamefont {Dadhich}}, \
  and\ \bibinfo {author} {\bibfnamefont {R.}~\bibnamefont {Maartens}},\ }\href
  {\doibase 10.1103/PhysRevD.65.101501} {\bibfield  {journal} {\bibinfo
  {journal} {Phys. Rev. D}\ }\textbf {\bibinfo {volume} {65}},\ \bibinfo
  {pages} {101501} (\bibinfo {year} {2002})}\BibitemShut {NoStop}%
\bibitem [{\citenamefont {Christodoulou}(1984)}]{Chr84}%
  \BibitemOpen
  \bibfield  {author} {\bibinfo {author} {\bibfnamefont {D.}~\bibnamefont
  {Christodoulou}},\ }\href {\doibase 10.1007/BF01223743} {\bibfield  {journal}
  {\bibinfo  {journal} {Comm. Math. Phys.}\ }\textbf {\bibinfo {volume} {93}},\
  \bibinfo {pages} {171} (\bibinfo {year} {1984})}\BibitemShut {NoStop}%
\bibitem [{\citenamefont {Eardley}\ and\ \citenamefont
  {Smarr}(1979)}]{PhysRevD.19.2239}%
  \BibitemOpen
  \bibfield  {author} {\bibinfo {author} {\bibfnamefont {D.~M.}\ \bibnamefont
  {Eardley}}\ and\ \bibinfo {author} {\bibfnamefont {L.}~\bibnamefont
  {Smarr}},\ }\href {\doibase 10.1103/PhysRevD.19.2239} {\bibfield  {journal}
  {\bibinfo  {journal} {Phys. Rev. D}\ }\textbf {\bibinfo {volume} {19}},\
  \bibinfo {pages} {2239} (\bibinfo {year} {1979})}\BibitemShut {NoStop}%
\bibitem [{\citenamefont {Waugh}\ and\ \citenamefont
  {Lake}(1988)}]{PhysRevD.38.1315}%
  \BibitemOpen
  \bibfield  {author} {\bibinfo {author} {\bibfnamefont {B.}~\bibnamefont
  {Waugh}}\ and\ \bibinfo {author} {\bibfnamefont {K.}~\bibnamefont {Lake}},\
  }\href {\doibase 10.1103/PhysRevD.38.1315} {\bibfield  {journal} {\bibinfo
  {journal} {Phys. Rev. D}\ }\textbf {\bibinfo {volume} {38}},\ \bibinfo
  {pages} {1315} (\bibinfo {year} {1988})}\BibitemShut {NoStop}%
\bibitem [{\citenamefont {Goswami}\ and\ \citenamefont
  {Joshi}(2007)}]{PhysRevD.76.084026}%
  \BibitemOpen
  \bibfield  {author} {\bibinfo {author} {\bibfnamefont {R.}~\bibnamefont
  {Goswami}}\ and\ \bibinfo {author} {\bibfnamefont {P.~S.}\ \bibnamefont
  {Joshi}},\ }\href {\doibase 10.1103/PhysRevD.76.084026} {\bibfield  {journal}
  {\bibinfo  {journal} {Phys. Rev. D}\ }\textbf {\bibinfo {volume} {76}},\
  \bibinfo {pages} {084026} (\bibinfo {year} {2007})}\BibitemShut {NoStop}%
\bibitem [{\citenamefont {Harada}\ \emph {et~al.}(1998)\citenamefont {Harada},
  \citenamefont {Iguchi},\ and\ \citenamefont {Nakao}}]{PhysRevD.58.041502}%
  \BibitemOpen
  \bibfield  {author} {\bibinfo {author} {\bibfnamefont {T.}~\bibnamefont
  {Harada}}, \bibinfo {author} {\bibfnamefont {H.}~\bibnamefont {Iguchi}}, \
  and\ \bibinfo {author} {\bibfnamefont {K.}~\bibnamefont {Nakao}},\ }\href
  {\doibase 10.1103/PhysRevD.58.041502} {\bibfield  {journal} {\bibinfo
  {journal} {Phys. Rev. D}\ }\textbf {\bibinfo {volume} {58}},\ \bibinfo
  {pages} {041502} (\bibinfo {year} {1998})}\BibitemShut {NoStop}%
\bibitem [{\citenamefont {Joshi}\ and\ \citenamefont
  {Dwivedi}(1993)}]{PhysRevD.47.5357}%
  \BibitemOpen
  \bibfield  {author} {\bibinfo {author} {\bibfnamefont {P.~S.}\ \bibnamefont
  {Joshi}}\ and\ \bibinfo {author} {\bibfnamefont {I.~H.}\ \bibnamefont
  {Dwivedi}},\ }\href {\doibase 10.1103/PhysRevD.47.5357} {\bibfield  {journal}
  {\bibinfo  {journal} {Phys. Rev. D}\ }\textbf {\bibinfo {volume} {47}},\
  \bibinfo {pages} {5357} (\bibinfo {year} {1993})}\BibitemShut {NoStop}%
\bibitem [{\citenamefont {Ori}\ and\ \citenamefont
  {Piran}(1987)}]{PhysRevLett.59.2137}%
  \BibitemOpen
  \bibfield  {author} {\bibinfo {author} {\bibfnamefont {A.}~\bibnamefont
  {Ori}}\ and\ \bibinfo {author} {\bibfnamefont {T.}~\bibnamefont {Piran}},\
  }\href {\doibase 10.1103/PhysRevLett.59.2137} {\bibfield  {journal} {\bibinfo
   {journal} {Phys. Rev. Lett.}\ }\textbf {\bibinfo {volume} {59}},\ \bibinfo
  {pages} {2137} (\bibinfo {year} {1987})}\BibitemShut {NoStop}%
\bibitem [{\citenamefont {Lake}(1991)}]{PhysRevD.43.1416}%
  \BibitemOpen
  \bibfield  {author} {\bibinfo {author} {\bibfnamefont {K.}~\bibnamefont
  {Lake}},\ }\href {\doibase 10.1103/PhysRevD.43.1416} {\bibfield  {journal}
  {\bibinfo  {journal} {Phys. Rev. D}\ }\textbf {\bibinfo {volume} {43}},\
  \bibinfo {pages} {1416} (\bibinfo {year} {1991})}\BibitemShut {NoStop}%
\bibitem [{\citenamefont {Shapiro}\ and\ \citenamefont
  {Teukolsky}(1991)}]{PhysRevLett.66.994}%
  \BibitemOpen
  \bibfield  {author} {\bibinfo {author} {\bibfnamefont {S.~L.}\ \bibnamefont
  {Shapiro}}\ and\ \bibinfo {author} {\bibfnamefont {S.~A.}\ \bibnamefont
  {Teukolsky}},\ }\href {\doibase 10.1103/PhysRevLett.66.994} {\bibfield
  {journal} {\bibinfo  {journal} {Phys. Rev. Lett.}\ }\textbf {\bibinfo
  {volume} {66}},\ \bibinfo {pages} {994} (\bibinfo {year} {1991})}\BibitemShut
  {NoStop}%
\bibitem [{\citenamefont {Stuchl{\'\i}k}\ and\ \citenamefont
  {Schee}(2010)}]{0264-9381-27-21-215017}%
  \BibitemOpen
  \bibfield  {author} {\bibinfo {author} {\bibfnamefont {Z.}~\bibnamefont
  {Stuchl{\'\i}k}}\ and\ \bibinfo {author} {\bibfnamefont {J.}~\bibnamefont
  {Schee}},\ }\href {http://stacks.iop.org/0264-9381/27/i=21/a=215017}
  {\bibfield  {journal} {\bibinfo  {journal} {Class. Quant. Grav.}\ }\textbf
  {\bibinfo {volume} {27}},\ \bibinfo {pages} {215017} (\bibinfo {year}
  {2010})}\BibitemShut {NoStop}%
\bibitem [{\citenamefont {Virbhadra}\ and\ \citenamefont
  {Keeton}(2008{\natexlab{a}})}]{PhysRevD.77.124014}%
  \BibitemOpen
  \bibfield  {author} {\bibinfo {author} {\bibfnamefont {K.~S.}\ \bibnamefont
  {Virbhadra}}\ and\ \bibinfo {author} {\bibfnamefont {C.~R.}\ \bibnamefont
  {Keeton}},\ }\href {\doibase 10.1103/PhysRevD.77.124014} {\bibfield
  {journal} {\bibinfo  {journal} {Phys. Rev. D}\ }\textbf {\bibinfo {volume}
  {77}},\ \bibinfo {pages} {124014} (\bibinfo {year}
  {2008}{\natexlab{a}})}\BibitemShut {NoStop}%
\bibitem [{\citenamefont {Bambi}\ and\ \citenamefont
  {Freese}(2009)}]{PhysRevD.79.043002}%
  \BibitemOpen
  \bibfield  {author} {\bibinfo {author} {\bibfnamefont {C.}~\bibnamefont
  {Bambi}}\ and\ \bibinfo {author} {\bibfnamefont {K.}~\bibnamefont {Freese}},\
  }\href {\doibase 10.1103/PhysRevD.79.043002} {\bibfield  {journal} {\bibinfo
  {journal} {Phys. Rev. D}\ }\textbf {\bibinfo {volume} {79}},\ \bibinfo
  {pages} {043002} (\bibinfo {year} {2009})}\BibitemShut {NoStop}%
\bibitem [{\citenamefont {Hioki}\ and\ \citenamefont
  {Maeda}(2009)}]{PhysRevD.80.024042}%
  \BibitemOpen
  \bibfield  {author} {\bibinfo {author} {\bibfnamefont {K.}~\bibnamefont
  {Hioki}}\ and\ \bibinfo {author} {\bibfnamefont {K.}~\bibnamefont {Maeda}},\
  }\href {\doibase 10.1103/PhysRevD.80.024042} {\bibfield  {journal} {\bibinfo
  {journal} {Phys. Rev. D}\ }\textbf {\bibinfo {volume} {80}},\ \bibinfo
  {pages} {024042} (\bibinfo {year} {2009})}\BibitemShut {NoStop}%
\bibitem [{\citenamefont {Bambi}\ \emph {et~al.}(2009)\citenamefont {Bambi},
  \citenamefont {Freese}, \citenamefont {Harada}, \citenamefont {Takahashi},\
  and\ \citenamefont {Yoshida}}]{PhysRevD.80.104023}%
  \BibitemOpen
  \bibfield  {author} {\bibinfo {author} {\bibfnamefont {C.}~\bibnamefont
  {Bambi}}, \bibinfo {author} {\bibfnamefont {K.}~\bibnamefont {Freese}},
  \bibinfo {author} {\bibfnamefont {T.}~\bibnamefont {Harada}}, \bibinfo
  {author} {\bibfnamefont {R.}~\bibnamefont {Takahashi}}, \ and\ \bibinfo
  {author} {\bibfnamefont {N.}~\bibnamefont {Yoshida}},\ }\href {\doibase
  10.1103/PhysRevD.80.104023} {\bibfield  {journal} {\bibinfo  {journal} {Phys.
  Rev. D}\ }\textbf {\bibinfo {volume} {80}},\ \bibinfo {pages} {104023}
  (\bibinfo {year} {2009})}\BibitemShut {NoStop}%
\bibitem [{\citenamefont {Bambi}\ \emph {et~al.}(2010)\citenamefont {Bambi},
  \citenamefont {Harada}, \citenamefont {Takahashi},\ and\ \citenamefont
  {Yoshida}}]{PhysRevD.81.104004}%
  \BibitemOpen
  \bibfield  {author} {\bibinfo {author} {\bibfnamefont {C.}~\bibnamefont
  {Bambi}}, \bibinfo {author} {\bibfnamefont {T.}~\bibnamefont {Harada}},
  \bibinfo {author} {\bibfnamefont {R.}~\bibnamefont {Takahashi}}, \ and\
  \bibinfo {author} {\bibfnamefont {N.}~\bibnamefont {Yoshida}},\ }\href
  {\doibase 10.1103/PhysRevD.81.104004} {\bibfield  {journal} {\bibinfo
  {journal} {Phys. Rev. D}\ }\textbf {\bibinfo {volume} {81}},\ \bibinfo
  {pages} {104004} (\bibinfo {year} {2010})}\BibitemShut {NoStop}%
\bibitem [{\citenamefont {Kov\'acs}\ and\ \citenamefont
  {Harko}(2010)}]{PhysRevD.82.124047}%
  \BibitemOpen
  \bibfield  {author} {\bibinfo {author} {\bibfnamefont {Z.}~\bibnamefont
  {Kov\'acs}}\ and\ \bibinfo {author} {\bibfnamefont {T.}~\bibnamefont
  {Harko}},\ }\href {\doibase 10.1103/PhysRevD.82.124047} {\bibfield  {journal}
  {\bibinfo  {journal} {Phys. Rev. D}\ }\textbf {\bibinfo {volume} {82}},\
  \bibinfo {pages} {124047} (\bibinfo {year} {2010})}\BibitemShut {NoStop}%
\bibitem [{\citenamefont {Pugliese}\ \emph
  {et~al.}(2011{\natexlab{c}})\citenamefont {Pugliese}, \citenamefont
  {Quevedo},\ and\ \citenamefont {Ruffini}}]{PhysRevD.83.024021}%
  \BibitemOpen
  \bibfield  {author} {\bibinfo {author} {\bibfnamefont {D.}~\bibnamefont
  {Pugliese}}, \bibinfo {author} {\bibfnamefont {H.}~\bibnamefont {Quevedo}}, \
  and\ \bibinfo {author} {\bibfnamefont {R.}~\bibnamefont {Ruffini}},\ }\href
  {\doibase 10.1103/PhysRevD.83.024021} {\bibfield  {journal} {\bibinfo
  {journal} {Phys. Rev. D}\ }\textbf {\bibinfo {volume} {83}},\ \bibinfo
  {pages} {024021} (\bibinfo {year} {2011}{\natexlab{c}})}\BibitemShut
  {NoStop}%
\bibitem [{\citenamefont {Janis}\ \emph {et~al.}(1968)\citenamefont {Janis},
  \citenamefont {Newman},\ and\ \citenamefont {Winicour}}]{PhysRevLett.20.878}%
  \BibitemOpen
  \bibfield  {author} {\bibinfo {author} {\bibfnamefont {A.~I.}\ \bibnamefont
  {Janis}}, \bibinfo {author} {\bibfnamefont {E.~T.}\ \bibnamefont {Newman}}, \
  and\ \bibinfo {author} {\bibfnamefont {J.}~\bibnamefont {Winicour}},\ }\href
  {\doibase 10.1103/PhysRevLett.20.878} {\bibfield  {journal} {\bibinfo
  {journal} {Phys. Rev. Lett.}\ }\textbf {\bibinfo {volume} {20}},\ \bibinfo
  {pages} {878} (\bibinfo {year} {1968})}\BibitemShut {NoStop}%
\bibitem [{\citenamefont {Patil}\ and\ \citenamefont
  {Joshi}(2012)}]{patil2012acceleration}%
  \BibitemOpen
  \bibfield  {author} {\bibinfo {author} {\bibfnamefont {M.}~\bibnamefont
  {Patil}}\ and\ \bibinfo {author} {\bibfnamefont {P.~S.}\ \bibnamefont
  {Joshi}},\ }\href@noop {} {\bibfield  {journal} {\bibinfo  {journal} {Phys.
  Rev. D}\ }\textbf {\bibinfo {volume} {85}},\ \bibinfo {pages} {104014}
  (\bibinfo {year} {2012})}\BibitemShut {NoStop}%
\bibitem [{\citenamefont {Kovacs}\ and\ \citenamefont
  {Harko}(2010)}]{kovacs2010can}%
  \BibitemOpen
  \bibfield  {author} {\bibinfo {author} {\bibfnamefont {Z.}~\bibnamefont
  {Kovacs}}\ and\ \bibinfo {author} {\bibfnamefont {T.}~\bibnamefont {Harko}},\
  }\href@noop {} {\bibfield  {journal} {\bibinfo  {journal} {Phys. Rev. D}\
  }\textbf {\bibinfo {volume} {82}},\ \bibinfo {pages} {124047} (\bibinfo
  {year} {2010})}\BibitemShut {NoStop}%
\bibitem [{\citenamefont {Bozza}(2002)}]{bozza2002gravitational}%
  \BibitemOpen
  \bibfield  {author} {\bibinfo {author} {\bibfnamefont {V.}~\bibnamefont
  {Bozza}},\ }\href@noop {} {\bibfield  {journal} {\bibinfo  {journal} {Phys.
  Rev. D}\ }\textbf {\bibinfo {volume} {66}},\ \bibinfo {pages} {103001}
  (\bibinfo {year} {2002})}\BibitemShut {NoStop}%
\bibitem [{\citenamefont {Virbhadra}\ and\ \citenamefont
  {Ellis}(2002)}]{PhysRevD.65.103004}%
  \BibitemOpen
  \bibfield  {author} {\bibinfo {author} {\bibfnamefont {K.~S.}\ \bibnamefont
  {Virbhadra}}\ and\ \bibinfo {author} {\bibfnamefont {G.~F.~R.}\ \bibnamefont
  {Ellis}},\ }\href {\doibase 10.1103/PhysRevD.65.103004} {\bibfield  {journal}
  {\bibinfo  {journal} {Phys. Rev. D}\ }\textbf {\bibinfo {volume} {65}},\
  \bibinfo {pages} {103004} (\bibinfo {year} {2002})}\BibitemShut {NoStop}%
\bibitem [{\citenamefont {Virbhadra}\ and\ \citenamefont
  {Keeton}(2008{\natexlab{b}})}]{virbhadra2008time}%
  \BibitemOpen
  \bibfield  {author} {\bibinfo {author} {\bibfnamefont {K.}~\bibnamefont
  {Virbhadra}}\ and\ \bibinfo {author} {\bibfnamefont {C.}~\bibnamefont
  {Keeton}},\ }\href@noop {} {\bibfield  {journal} {\bibinfo  {journal} {Phys.
  Rev. D}\ }\textbf {\bibinfo {volume} {77}},\ \bibinfo {pages} {124014}
  (\bibinfo {year} {2008}{\natexlab{b}})}\BibitemShut {NoStop}%
\bibitem [{\citenamefont {Gyulchev}\ and\ \citenamefont
  {Yazadjiev}(2008)}]{PhysRevD.78.083004}%
  \BibitemOpen
  \bibfield  {author} {\bibinfo {author} {\bibfnamefont {G.~N.}\ \bibnamefont
  {Gyulchev}}\ and\ \bibinfo {author} {\bibfnamefont {S.~S.}\ \bibnamefont
  {Yazadjiev}},\ }\href {\doibase 10.1103/PhysRevD.78.083004} {\bibfield
  {journal} {\bibinfo  {journal} {Phys. Rev. D}\ }\textbf {\bibinfo {volume}
  {78}},\ \bibinfo {pages} {083004} (\bibinfo {year} {2008})}\BibitemShut
  {NoStop}%
\bibitem [{\citenamefont {Liao}\ \emph {et~al.}(2014)\citenamefont {Liao},
  \citenamefont {Chen}, \citenamefont {Huang},\ and\ \citenamefont
  {Wang}}]{LiaCheHua14}%
  \BibitemOpen
  \bibfield  {author} {\bibinfo {author} {\bibfnamefont {P.}~\bibnamefont
  {Liao}}, \bibinfo {author} {\bibfnamefont {J.~H.}\ \bibnamefont {Chen}},
  \bibinfo {author} {\bibfnamefont {H.}~\bibnamefont {Huang}}, \ and\ \bibinfo
  {author} {\bibfnamefont {Y.~J.}\ \bibnamefont {Wang}},\ }\href
  {http://dx.doi.org/10.1007/s10714-014-1752-9} {\bibfield  {journal} {\bibinfo
   {journal} {Gen. Rel. Grav.}\ }\textbf {\bibinfo {volume} {46}},\ \bibinfo
  {eid} {1752} (\bibinfo {year} {2014})}\BibitemShut {NoStop}%
\bibitem [{\citenamefont {Chowdhury}\ \emph {et~al.}(2012)\citenamefont
  {Chowdhury}, \citenamefont {Patil}, \citenamefont {Malafarina},\ and\
  \citenamefont {Joshi}}]{PhysRevD.85.104031}%
  \BibitemOpen
  \bibfield  {author} {\bibinfo {author} {\bibfnamefont {A.~N.}\ \bibnamefont
  {Chowdhury}}, \bibinfo {author} {\bibfnamefont {M.}~\bibnamefont {Patil}},
  \bibinfo {author} {\bibfnamefont {D.}~\bibnamefont {Malafarina}}, \ and\
  \bibinfo {author} {\bibfnamefont {P.~S.}\ \bibnamefont {Joshi}},\ }\href
  {\doibase 10.1103/PhysRevD.85.104031} {\bibfield  {journal} {\bibinfo
  {journal} {Phys. Rev. D}\ }\textbf {\bibinfo {volume} {85}},\ \bibinfo
  {pages} {104031} (\bibinfo {year} {2012})}\BibitemShut {NoStop}%
\end{thebibliography}%

\end{document}